\documentclass[11pt]{article}
\usepackage{latexsym}
\usepackage{amssymb}
\usepackage{epsf}
\usepackage{amsmath}
\usepackage[hypertex]{hyperref}
\usepackage{graphicx}% Include figure files

\newcommand{\lsim}{\lesssim}
\newcommand{\gsim}{\gtrsim}

\newcommand{\beq}{\begin{equation}}   
\newcommand{\eeq}{\end{equation}}
\newcommand{\bea}{\begin{eqnarray}}   
\newcommand{\eea}{\end{eqnarray}}
\newcommand{\bear}{\begin{array}}  
\newcommand {\eear}{\end{array}}
\newcommand{\bef}{\begin{figure}}  
\newcommand {\eef}{\end{figure}}
\newcommand{\bec}{\begin{center}}  
\newcommand {\eec}{\end{center}}
\newcommand{\non}{\nonumber}  

\newcommand{\la}{\left\langle}  
\newcommand{\ra}{\right\rangle}
\newcommand{\ds}{\displaystyle}

\def\EQ#1{Eq.~(\ref{#1})}

\def\REF#1{(\ref{#1})}
\def\GEV#1{10^{#1}{\rm\,GeV}}

\def\lrf#1#2{ \left(\frac{#1}{#2}\right)}
\def\lrfp#1#2#3{ \left(\frac{#1}{#2} \right)^{#3}}
\def\vev#1{\langle #1\rangle} % added by Hama

\begin{document}
\baselineskip=18pt

\begin{titlepage}

\begin{flushright}
UT-09-17\\
IPMU09-0088\\
TU-850
\end{flushright}

\vskip 1.35cm
\begin{center}
{\Large \bf
Non-thermal Gravitino Dark Matter in Gauge Mediation\\
}
\vskip 1.2cm
Koichi Hamaguchi$^{1,2}$, Ryuichiro Kitano$^{3}$, Fuminobu Takahashi$^2$
\vskip 0.4cm

{\it $^1$  Department of Physics, University of Tokyo,
  Tokyo 113-0033, Japan\\
$^2$ Institute for the Physics and Mathematics of the Universe, 
University of Tokyo,\\ Chiba 277-8568, Japan\\
$^3$
Department of Physics, Tohoku University, Sendai 980-8578, Japan
}

\vskip 1.5cm

\abstract{ 
  We show that gravitinos produced by decays of a supersymmetry
  breaking scalar field (the pseudo-moduli field) can naturally
  explain the observed abundance of dark matter in a certain class of
  the gauge mediation models. We study the decay processes as well as
  cosmological constraints on this scenario in detail, particularly
  focusing on different behavior of the real and imaginary components
  of the pseudo-moduli field. Cosmologically viable scenario emerges
  when the gravitino and the pseudo-moduli masses are ${\cal
    O}(10-100)$~MeV and ${\cal O}(100)$~GeV, respectively.
}
\end{center}
\end{titlepage}

\setcounter{footnote}{0}
\setcounter{page}{2}

%%%%%%%%%%%%%%%%%%%%%%%%%%%%%%%%%%%%
\section{Introduction}
%%%%%%%%%%%%%%%%%%%%%%%%%%%%%%%%%%%%
The presence of dark matter (DM) in the Universe was firmly
established by numerous
observations~\cite{Komatsu:2008hk}. Nevertheless, it has remained as a
big mystery in cosmology as well as particle physics what DM is made
of.  Since there is no candidate for DM in the standard model, we need
to consider new physics.

In a supersymmetric extension of the standard model, the lightest
supersymmetric particle (LSP) is a stable particle if the $R$ parity
is conserved. Depending on the mediation mechanism of the
supersymmetry (SUSY) breaking effects, the lightest neutralino or the
gravitino can thus be a good candidate for DM. The latter possibility
is naturally realized in a framework of gauge
mediation~\cite{Dine:1981za,Dine:1993yw}, which has a virtue of
avoiding the SUSY flavor problem.

The production mechanisms of gravitinos in the early Universe are
broadly classified into thermal or non-thermal one. The thermal
production is always present as long as the Universe becomes radiation
dominated after
inflation~\cite{Weinberg:1982zq,Krauss:1983ik,Moroi:1993mb,deGouvea:1997tn,
Bolz:1998ek,Bolz:2000fu,Pradler:2006qh}. In this case the decay rate of
an inflaton must be such that gravitinos, produced by particle
scatterings in thermal plasma, account for the observed DM abundance. If
the inflationary dynamics has nothing to do with the SUSY breaking
mechanism (and therefore the gravitino mass), such a coincidence may
call for some explanation.  On the other hand, non-thermal gravitino
production has been discussed (mostly as a problem of overproduction or
a solution to it) in the context of the decay of the next to lightest
SUSY particle (NLSP)~\cite{Feng:2003xh,Feng:2004mt}, the
moduli~\cite{Hashimoto:1998mu,
Moroi:1999zb,Endo:2006zj,Nakamura:2006uc,Dine:2006ii,Endo:2006tf}, the
inflaton~\cite{Felder:1999pv,Maroto:1999ch,Kallosh:1999jj,
Giudice:1999yt,
Kawasaki:2006gs,Asaka:2006bv,Endo:2007ih,Takahashi:2007tz} and the SUSY
breaking field (sometimes called as the Polonyi field or the
pseudo-moduli field)~\cite{Dine:1983ys, Coughlan:1984yk, Banks:1993en,
Joichi:1994ce, Ibe:2006am,Ibe:2006rc,Heckman:2008jy}.  In particular, it
is interesting to see if a right amount of the gravitinos can be
produced by the decay of the SUSY breaking field, since the structure of
the SUSY breaking sector may be probed by cosmological arguments.

In this paper, we investigate the gravitino DM scenario in a generic
setup of gauge-mediated SUSY breaking models.  In many SUSY breaking
models, there is a light singlet scalar field, which obtains a mass
from SUSY breaking.  During the inflation era, this scalar field, the
pseudo-moduli field, can have a large displacement from the true
vacuum, and at a later time, it starts coherent oscillations about the
minimum of the potential. Under reasonable assumptions, the
oscillation energy dominates the energy density of the Universe, and
the decay of the scalar field produces radiation as well as gravitinos
which remain as DM today.  We update the calculation of
Ref.~\cite{Ibe:2006rc} by taking into account the following points. We
do not assume a particular relation among parameters in the SUSY
breaking sector.  We treat the real and the imaginary parts of the
scalar field separately as their decay properties are quite different.
We find that, for the mass of the scalar field around ${\cal
  O}(100)$~GeV and the gravitino mass of ${\cal O}(10-100)$~MeV, the
decay of the imaginary part is the main source of the radiation and
the gravitinos that account for the observed DM abundance.  The region
turns out to be similar to the one found in Ref.~\cite{Ibe:2006rc}
although the main decay mode is different.  The consistent region
overlaps with the prediction of a model in Ref.~\cite{Ibe:2007km}
where the $\mu$ problem is solved.\footnote{A similar parameter region
  is identified in Ref.~\cite{Heckman:2008jy} in the F-theory GUT
  model by considering the abundance of gravitino dark matter and the
  $\mu$-problem. In discussions of cosmology, the most important
  difference between two models is the mass of the imaginary part of
  the SUSY breaking scalar field. In the F-theory GUT model, the
  imaginary part is assumed to be much lighter than the real part. In
  the model of Refs.~\cite{Kitano:2006wz, Ibe:2006rc,Ibe:2007km}, in
  contrast, the real and the imaginary parts have almost the same
  masses.  }

%%%%%%%%%%%%%%%%%%%%%%%%%%%%%%%%%%%%%%%%%
\section{Gauge mediation}
%%%%%%%%%%%%%%%%%%%%%%%%%%%%%%%%%%%%%%%%%

We first define the framework and identify parameters relevant for the
discussion of cosmology.  We use an effective description of
gauge-mediation models given in terms of a SUSY breaking field $S$ and
the fields in the minimal SUSY standard model (MSSM).  The SUSY
breaking sector is described by a single chiral superfield $S$ which
consists of the Goldstino fermion and its scalar partner $s$ with the
K\"ahler and super-potentials:
\begin{eqnarray}
K = S^\dagger S - {(S^\dagger S)^2 \over \Lambda^2}+ ({\mbox{higher
order}}),
\label{eq:kahler}
\end{eqnarray}
\begin{eqnarray}
W = m^2 S,
\label{eq:super}
\end{eqnarray}
where $\Lambda$ is a cut-off scale of the effective theory, and $m$
denotes the size of the SUSY breaking.  This form is obtained after
integrating out massive fields in a wide class of SUSY breaking
models.
The equation of motion gives $F_S = m^2$ as long as there is no
singularity in the K{\" a}hler potential. The second term in
Eq.~(\ref{eq:kahler}) stabilizes the scalar potential at $s=0$.

The MSSM particles can couple to the SUSY breaking sector through
messenger fields, $f$ and $\bar f$:
\begin{eqnarray}
W \ni - \lambda S f \bar f,
\label{eq:messenger}
\end{eqnarray}
where $\lambda$ is a dimensionless coupling constant. 
With this term, the potential is minimized at $s=0$ and $f \bar f =
m^2 /\lambda$ where SUSY is unbroken and the gauge symmetry of the
MSSM is broken.
Therefore, one needs some mechanism to stabilize the potential at
$\langle s \rangle \neq 0$ and $\langle f \rangle = \langle \bar f
\rangle = 0$.

In order to keep discussion as general as possible, we do not specify
such a mechanism in the following and treat the three quantities
($\langle s \rangle$, $F_S$, $\Lambda$) as independent
parameters. Here we define the origin of the $s$ field to be the point
where the messenger fields become massless, i.e., $M_{\rm mess} =
\lambda \langle s \rangle$. Once we integrate out the messenger
fields, the gauge kinetic term in this case is given by
\begin{eqnarray}
f = {1 \over 2} 
\left(
{1 \over g^2} -
{2 N \over (4 \pi)^2} \log {S \over \Lambda} 
\right)
W^\alpha W_\alpha + {\rm h.c},
\label{eq:gauge_kf}
\end{eqnarray}
where $g$ is the gauge coupling constant and $N$ is the (effective)
number of messenger fields.
Our later discussion can apply when the low energy effective theory is
of this type.
For example, in the model of Ref.~\cite{Kitano:2006wz} the
supergravity effects create a local minimum at $\langle s \rangle \sim
\Lambda^2 / M_{\rm Pl}$, where $M_{\rm Pl} \simeq 2.4 \times
10^{18}$\,GeV is the reduced Planck scale.
Ref.~\cite{Murayama:2006yf} discussed a model with an additional
superpotential term, $W \ni M_{\rm mess} f \bar f$, with which the
effective value of $\langle s \rangle$ is $M_{\rm mess} / \lambda$.

The important point here is that the scalar field $s$ couples to the
MSSM fields with a suppression of $F_S/\langle s \rangle^2$, whereas
the coupling to the gravitino is suppressed by
$F_S/\Lambda^2$. Therefore, for $\langle s \rangle \ll \Lambda$, there
is a possibility to avoid the dangerous gravitino overproduction as
well as the catastrophic entropy release from the $s$
decay~\cite{Coughlan:1983ci, Dine:1983ys, Banks:1993en, Joichi:1994ce,
  Endo:2006zj, Nakamura:2006uc, Dine:2006ii, Ibe:2006am}.

The three parameters $\langle s \rangle$, $F_S$, and $\Lambda$ can be
expressed in terms of physical quantities relevant for our discussion,
such as the masses of Bino, $s$, and gravitino [$m_{\tilde B}$, $m_S$,
$m_{3/2}$].  They are related to $\langle s \rangle$, $F_S$, and
$\Lambda$ as
\bea
\label{eq:gaugino}
m_{\tilde B} &=& {g_1^2 N \over (4 \pi)^2} {F_S \over \langle s \rangle},\\
\label{eq:mS}
m_S &=& {2 F_S \over \Lambda},\\
m_{3/2} &=& {F_S \over \sqrt{3} M_{\rm Pl}},
\label{eq:gravitino}
\end{eqnarray}
where $g_1 = \sqrt{5/3} g_Y$ with $g_Y$ being the coupling constant of
the U(1)$_Y$ gauge interaction.
When expressed in terms of the running Bino mass at the electroweak
scale, explicit $\lambda$ dependence of the low energy quantities
disappears in many places.
We can invert the above relations and write $\langle s \rangle$,
$F_S$, and $\Lambda$ in terms of the physical quantities:
\begin{eqnarray}
F_S &=& 1.3\times \GEV{17}^2 
\lrf{m_{3/2}}{30~{\rm MeV}},
\\
\vev{s} &=& 8.6 \times \GEV{11}
\cdot N \,
\lrf{m_{3/2}}{30~{\rm MeV}}
\lrfp{m_{\tilde B}}{200~{\rm GeV}}{-1},
\label{eq:svev}\\
\Lambda &=& 2.5 \times \GEV{15}
\lrf{m_{3/2}}{30~{\rm MeV}}
\lrfp{m_S}{100~{\rm GeV}}{-1}.
\end{eqnarray}
Here and in what follows, we use $m_{3/2}=30~{\rm MeV}$, $m_{\tilde
  B}=200~{\rm GeV}$ and $m_S=100~{\rm GeV}$ as reference values,
though the following discussion is generic and does not depend on
those explicit values.

Although $\lambda$ does not appear in the above relations, it cannot
take an arbitrary value. In fact, there are lower and upper bounds on
$\lambda$ to avoid instabilities at the SUSY breaking minimum. In
order to avoid a tachyonic mass for the messenger fields, $\lambda$
should satisfy $\lambda^2 \vev{s}^2 > \lambda F_S$, i.e.,
\beq
\lambda > \lambda_{\rm min} = 1.7\times 10^{-7}\cdot N^{-2}
\lrfp{m_{3/2}}{30~{\rm MeV}}{-1}
\lrfp{m_{\tilde B}}{200~{\rm GeV}}{2}.
\eeq
On the other hand, the interaction term in \EQ{eq:messenger} induces a
logarithmic potential at one-loop level~\cite{Ibe:2006rc,Kitano:2006wz},
\beq
V(s) = m_S^2 |s-\vev{s}|^2 
+ \frac{5N}{16\pi^2}\lambda^2 F_S^2\log\lrf{|s|^2}{\Lambda^2},
\label{eq:Vlog}
\eeq
where we have assumed that the messenger fields transform as ${\bf 5}$
and $\bf{\bar 5}$ under SU(5). The logarithmic potential gives an
attractive force on the $s$ field toward the SUSY vacuum at the
origin. The stability at the SUSY breaking minimum requires
\beq
\frac{5N}{16\pi^2}\lambda^2 F_S^2 < \frac{1}{4}m_S^2\vev{s}^2,
\eeq
namely
\beq
\lambda < \lambda_{\rm max} = 1.9\times 10^{-3}\cdot N^{1/2}
\lrf{m_S}{100~{\rm GeV}}
\lrfp{m_{\tilde B}}{200~{\rm GeV}}{-1}.
\eeq
In the following, we assume $\lambda_{\rm min} < \lambda < \lambda_{\rm
max}$.  The above logarithmic potential also induces a mass splitting
between the real and imaginary parts of the $s$ field, $\delta
m^2/m_S^2= {\cal O}(\lambda^2/\lambda_{\rm max}^2)$, as well as a shift
of the minimum, $\delta \vev{s}/\vev{s} = {\cal
O}(\lambda^2/\lambda_{\rm max}^2)$.  For simplicity, we assume
$\lambda^2 \ll \lambda_{\rm max}^2$ and neglect those corrections in the
following discussion.

%

%%%%%%%%%%%%%%%%%%%%%%%%%%%%%%%%%%%%
\section{Scenario}
%%%%%%%%%%%%%%%%%%%%%%%%%%%%%%%%%%%%

Let us first give an overview of the cosmological scenario in this
model; (i) the $s$ field develops a large expectation value during the
inflation, (ii) its coherent oscillations after the inflation dominate
the energy density of the Universe, and (iii) its decay produces
radiation (including SUSY particles if kinematically allowed) and
gravitinos.  The assumption (i) is quite natural as far as $m_S$ is
much smaller than the Hubble parameter $H$ during the inflation, since
the minimum of the potential during and after the inflation can be
well separated from $s = 0$ due to the deformation of the potential
through gravitational (or general $1/M_{\rm Pl}$ suppressed)
interactions.  The $s$ field then starts to oscillate around the
minimum when the Hubble parameter becomes comparable to $m_S$ and
keeps oscillating until it decays.

\begin{figure}[t]
\begin{center}
\includegraphics[width=10cm]{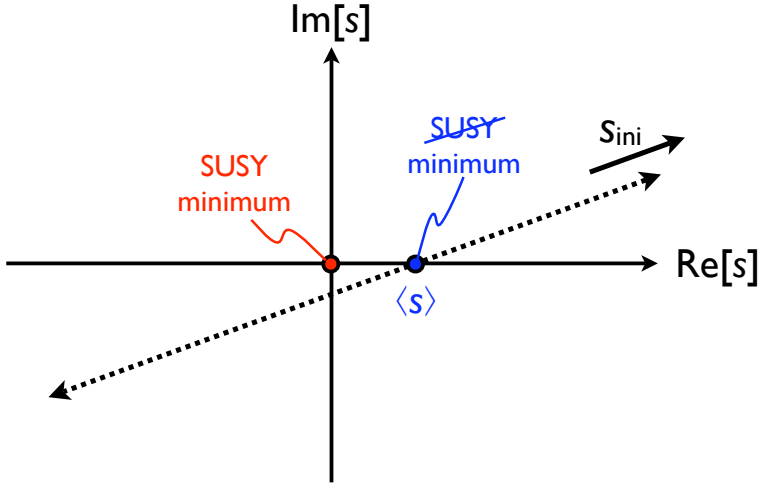}
\end{center}
\caption{A schematic figure of the evolution of $s$ field.}
\label{fig:sRsI}
\end{figure}

In the rest of this section, we discuss several conditions for the
above scenario to work.  An important point here is that there is a
global SUSY minimum of the potential at $s=0$, apart from the local
SUSY breaking minimum, $s=\vev{s}$.  Hereafter, we take the basis
where $\vev{s}$ is real.  (See Fig.~\ref{fig:sRsI}.)  As discussed in
Ref.~\cite{Ibe:2006rc}, the $s$ field does not fall into the SUSY
vacuum unless its initial value $s_{\rm ini}$ is too close to the real
axis.  We investigate in more detail the conditions for $s$ to be
trapped at the SUSY breaking minimum.
When discussing the dynamics of the $s$ field, we neglect the
corrections from higher order terms in the K\"ahler potential in
\EQ{eq:kahler}, which is small as far as $|s| \lsim \Lambda$.

%%%figure%%%
%%%figure%%%

%%%%%%%%%%%%%%%%%%%%%%%%%%
\subsection{Avoiding the SUSY vacuum}
%%%%%%%%%%%%%%%%%%%%%%%%%%
If the value of $s$ approaches too close to $s=0$ during the
oscillations, the scalar components of the messenger fields may become
tachyonic, which makes the Universe quickly fall into the SUSY vacuum.
This can be avoided if the initial value for ${\rm Im} [s]$ is so
large that the trajectory of $s$ stays away from $s=0$, satisfying
\begin{eqnarray}
|s| > \sqrt{ F_S \over \lambda }
\label{eq:region1}
\end{eqnarray}
in the course of oscillations.

Even if the above condition is met, the motion of $s$ can be
significantly affected by the deformation of the potential near the
origin due to the logarithmic potential \EQ{eq:Vlog}.  Moreover, it is
known that a scalar field oscillating on a scalar potential of the
logarithmic form experiences strong spatial instabilities and quickly
deforms into spatially random and inhomogeneous
state~\cite{Kusenko:1997si}.  If such instabilities becomes
significant before the $s$ field gets trapped in the SUSY breaking
minimum, we expect that it falls into the SUSY vacuum. This can be
avoided if the logarithmic correction remains subdominant along the
trajectory passing near the origin, i.e., $|s| \ll \la s \ra$, and the
condition is given by
\beq
|s|^2 \;\gtrsim\; \frac{5N}{16\pi^2} \frac{\lambda^2 F_S^2}{m_S^2}.
\label{eq:log}
\eeq
More rigorous derivation of \REF{eq:log} can be found in Appendix \ref{app:si}.

The minimum value of $|s|$ during the oscillations is approximately
given by (cf. Fig.~\ref{fig:sRsI})
\begin{eqnarray}
|s|_{\rm min} &\simeq& \vev{s} \frac{{\rm Im}[s_{\rm ini}]}{|s_{\rm ini}|}\,,
\end{eqnarray}
where $s_{\rm ini}$ is the value of $s$ when it starts coherent
oscillations at $H \sim m_S$.  We can rewrite the constraints
(\ref{eq:region1}) and (\ref{eq:log}) respectively in terms of the
ratio of the initial amplitudes, $r \equiv \left| {{\rm Im}[s_{\rm
      ini}] / {\rm Re}[s_{\rm ini}]} \right|$, as
\begin{eqnarray}
\label{eq:r-bound1}
\frac{r}{\sqrt{1+r^2}} &\gtrsim& 0.013 \cdot
N^{-1} 
\left(
\lambda \over 10^{-3}
\right)^{-1/2}
\left(
m_{\tilde B} \over 200~{\rm GeV}
\right)
\lrfp{ m_{3/2}}{ 30~{\rm MeV}}{-1/2}, \\
\frac{r}{\sqrt{1+r^2}} &\gtrsim& 0.26 \cdot N^{-1/2}
\lrf{\lambda}{10^{-3}}
\lrf{m_{\tilde B} }{ 200~{\rm GeV}}
\lrfp{m_S}{100{\rm\,GeV}}{-1}.
\label{eq:r-bound2}
\end{eqnarray}

\begin{figure}[t]
\begin{center}
\includegraphics[width=8cm]{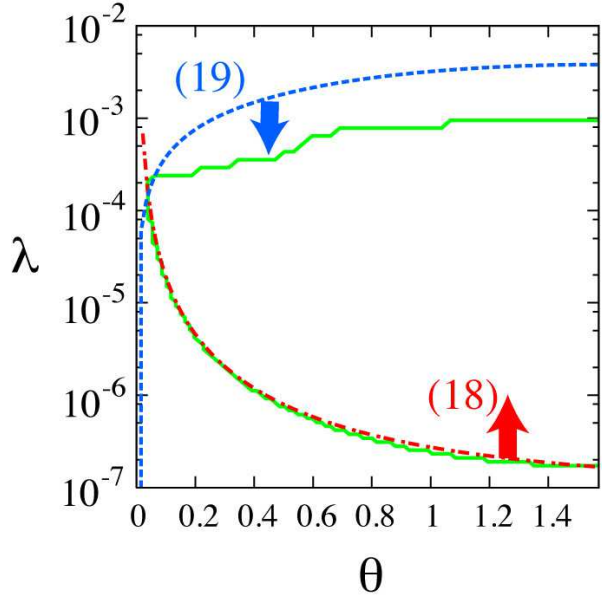}
\end{center}
\caption{The region surrounded by the solid (green) line represents a
  parameter space where the perturbation $\delta s$ remains smaller
  than the homogeneous part ${\bar s}$ until the $s$ field gets
  stabilized at the SUSY breaking minimum.  The conditions
  (\ref{eq:r-bound1}) and (\ref{eq:r-bound2}) are also shown as the
  dot-dashed (red) and dashed (blue) lines. We have chosen
  $m_{3/2}=30~{\rm MeV}$, $m_{\tilde B}=200~{\rm GeV}$,
  $m_S=100~{\rm GeV}$, and $N=1$.  }
\label{fig:allowed}
\end{figure}

In order to study the spatial instabilities, we separate the $s$ field
into a homogeneous part ${\bar s}$ and a perturbation $\delta s$.  We
have numerically followed the evolution of ${\bar s}$ and $\delta s$
for a set of reference values of $m_{3/2}$, $m_{\tilde B}$, $m_S$, and $N$,
keeping only terms linear in $\delta s$.  The initial conditions are
set as $ {\rm Re}[s_{\rm ini}] = \sqrt{2}\Lambda \cos \theta$ and
${\rm Im}[s_{\rm ini}] = \sqrt{2}\,\Lambda \sin \theta$, where
$\theta$ is related to $r$ as $\tan \theta = r$. We have chosen the
initial value of $|\delta s|$ equal to $10^{-5} \Lambda$, but the
following result is not sensitive to this value. (Indeed, we have
confirmed that the result remains almost intact for $|\delta s| =
10^{-10} \Lambda$.)  In Fig.~\ref{fig:allowed} we show a parameter
region surrounded by a solid (green) where $\delta s$ remains smaller
than ${\bar s}$ until the $s$ field settles down in the SUSY breaking
minimum. The allowed regions are found to be $10^{-7} \lesssim \lambda
\lesssim 10^{-3}$ and $0.03 \lesssim \theta \leq \pi/2$. For different
values of $m_{3/2}$, $m_{\tilde B}$ and $m_S$, the allowed regions are
modified correspondingly; in particular, the minimum value of $\theta$
can be smaller.  Note that $\delta s \sim {\bar s}$ does not
necessarily mean that the $s$ field falls into the supersymmetric
vacuum.  What we would like to emphasize here is that there {\it is} a
parameter space where our scenario is realized.  We have also shown in
the figure the conditions (\ref{eq:r-bound1}) and (\ref{eq:r-bound2}),
and the latter gives a slightly milder constraint on $\lambda$ than
the solid (green) line.  

Another concern is whether the ratio of the
energy densities of ${\rm Im}[s]$ and ${\rm Re}[s]$, $r_{\rm eff}^2
\equiv \rho_{s_I}/\rho_{s_R}$, is conserved or not.  According to our
numerical calculations, the final value of $r_{\rm eff}$ is always
larger or equal to $r$, and $r_{\rm eff}$ tends to become larger for
smaller $\theta$ and larger $\lambda$.  For most of the region
surrounded by the solid (green) line, however, the ratio does not
significantly evolve, and in particular, it remains almost constant in
the course of evolution for $\theta > 0.1$ and $\lambda < 10^{-4}$.
Therefore we do not distinguish $r_{\rm eff}$ from $r$ in the
following discussion. In the above analysis we have not taken into
consideration thermal effects, which will be discussed in Appendix
\ref{app:finiteT}.

%%%figure%%%
%%%figure%%%

%%%%%
\subsection{$s$-dominated Universe}

We assume that the initial value of $|s|$ is so large that the
coherent oscillations of $s$ dominate over the energy density of the
Universe before the time it decays.  Such a domination happens if
\begin{eqnarray}
T_d^{(R)} &<& 
{\rm min}(T_R, T_{m_S})
\left(
| {\rm Re}[s_{\rm ini}] - \vev{s} | \over \sqrt 3 M_{\rm Pl}
\right)^2
\label{eq:sRdomination}
\\{\rm or}~~~&&\non\\ 
T_d^{(I)} &<& 
{\rm min}(T_R, T_{m_S})
\left(
| {\rm Im}[s_{\rm ini}] | \over \sqrt 3 M_{\rm Pl}
\right)^2
\label{eq:sIdomination}
\end{eqnarray}
where $T_d^{(R/I)}$ are the decay temperatures of $s_{R/I}$ (cf. next
section), $T_R$ is the reheating temperature after the inflaton, and
$T_{m_S}$ is the temperature at $H=m_S$ in the radiation dominated
Universe,
\begin{eqnarray}
 T_{m_S} \;\simeq\; 7 \times 10^9~{\rm GeV} 
\left( {g_* \over 200} \right)^{-1/4}
\left( {m_S \over 100~{\rm GeV}}\right)^{1/2},
\end{eqnarray}
where $g_*$ counts the relativistic degrees of freedom in plasma.

%%%figure%%%
\begin{figure}[t]
\begin{center}
\includegraphics[width=12cm]{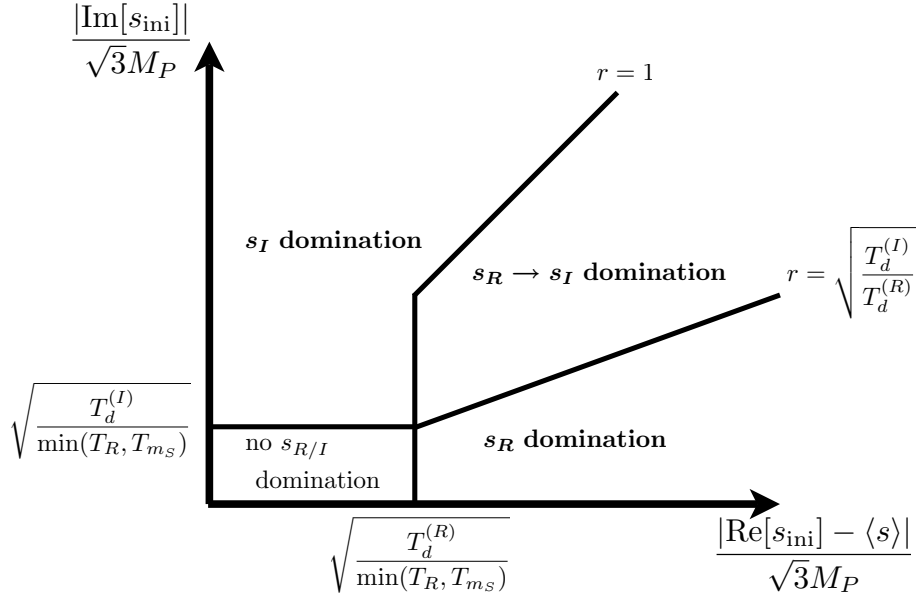}
\end{center}
\caption{The condition of the $s$-dominated Universe.}
\label{fig:s-dom}
\end{figure}
%%%figure%%%

As we will see in the next section, $T_d^{(I)}$ is always smaller than
or equal to $T_d^{(R)}$.  The history of the Universe depends on the
values of the decay temperatures and $r$.  There are four
possibilities (cf Fig.~\ref{fig:s-dom}):

{\bf Case 1:~[$s_I$-domination]} For $r>1$, the energy density of
$s_R$ is always smaller than that of $s_I$, irrespective of whether
(\ref{eq:sRdomination}) is satisfied or not.  The $s_I$ dominates the
energy of the Universe, if (\ref{eq:sIdomination}) is satisfied.

{\bf Case 2:~[$s_I$-domination after $s_R$-domination]} If
(\ref{eq:sRdomination}) is satisfied for $ \sqrt{T_d^{(I)}/
  T_d^{(R)}}<r<1$, it is $s_R$ that dominates the energy of the
Universe first, and the $s_I$ dominates the energy of the Universe
after the decay of $s_R$.

{\bf Case 3:~[$s_R$-domination]} If (\ref{eq:sRdomination}) is
satisfied for $r < \sqrt{T_d^{(I)}/ T_d^{(R)}}(<1)$, $s_R$ dominates
the energy of the Universe, while $s_I$ does not.

{\bf Case 4:~[$s_I$-domination]} If (\ref{eq:sRdomination}) is not
satisfied while (\ref{eq:sIdomination}) is satisfied for $r<1$, $s_I$
dominates the energy density of the Universe.

Once the energy density of the Universe is dominated by either $s_R$
or $s_I$, later discussions of the non-thermal gravitino production do
not depend on the reheating temperature $T_R$.  In Appendix
\ref{app:T}, we will consider the thermal effects, and a consistent
parameter region for the $s$ domination is found to be $10^{-5}
\lesssim \lambda \lesssim 10^{-3}$, $0.2 \lesssim \theta \leq \pi /
2$, and $10^5~{\rm GeV} \lesssim T_R \lesssim 10^6~{\rm GeV}$ for the
set of reference values: $m_{3/2}=30~{\rm MeV}$, $m_{\tilde
  B}=200~{\rm GeV}$, $m_S=100~{\rm GeV}$ and $N=1$.  We would like to
emphasize here that the consistent ranges for $\lambda$ and $\theta$
depend on the choice of $m_{3/2}$, $m_{\tilde B}$ and $m_S$. For
instance, the lowest allowed value of $\theta$ can be as small as
${\cal O}(0.01)$.

%%%%%%%%%%%%%%%%%%%%%%%%%%%%%%%%%%%%%%%
\section{Decays of the {\boldmath $s$} field}
%%%%%%%%%%%%%%%%%%%%%%%%%%%%%%%%%%%%%%%

The $s$ field mainly decays into the MSSM particles through loop
diagrams of the messenger fields. We first discuss the main decay mode
and calculate the decay temperatures.

\subsection{Decays of {\boldmath $s_R$}}

The effective couplings between $s_R$ and the MSSM fields can be read
off from the $\langle s \rangle$ dependencies of low-energy
parameters.
For scalar fields, the interaction terms are given by
\begin{eqnarray}
{\cal L}_{\rm int}^{(\tilde f)} 
= { \sqrt 2 (m_{\rm eff}^{(\tilde f)})^2  \over | \langle s \rangle | } \cdot
s_R | \tilde f |^2.
\label{eq:scalar}
\end{eqnarray}
The effective mass parameter $(m_{\rm eff}^{(\tilde f)})^2$ is a part
of the scalar mass that is proportional to $1/|\langle s \rangle|^2$,
i.e., $(m_{\rm eff}^{(\tilde f)})^2 = - d m_{\tilde f}^2 / d \log
|\langle s \rangle|^2$. If gauge mediation is the only contribution to
the scalar masses, $m_{\rm eff}^{(\tilde f)}$ is identical to their
masses. In realistic models of gauge mediation, the $\mu$ parameter
needs to be generated by some mechanism, and such contributions to the
masses of the Higgs fields may be independent of $\langle s \rangle$.

The couplings to the gauginos, $\lambda$, are
\begin{eqnarray}
{\cal L}_{\rm int}^{(\lambda)}
= { m_\lambda \over \sqrt 2 \langle s \rangle } \cdot
{ 1 \over 2}s_R \bar \lambda \lambda ,
\end{eqnarray}
where $m_\lambda$ is the gaugino mass (cf. \EQ{eq:gauge_kf}). There is a similar coupling
between $s_R$ to Higgsinos: 
\begin{eqnarray}
{\cal L}_{\rm int}^{(\rm Higgsino)} =
- {\mu_{\rm eff} \over \sqrt 2 \langle s \rangle }
 \cdot
 s_R
 \left(
 \overline{h_d^c} \cdot P_L h_u
 \right)
 + {\rm h.c.}
\end{eqnarray}
The coefficient $\mu_{\rm eff}$ is again a part of $\mu$ that is
proportional to $1/\langle s \rangle$.

There are couplings to the quarks and leptons through a mixing between
$s_R$ and Higgs bosons ($h^0$ and $H^0$). The mixing is induced
through the interaction term in Eq.~(\ref{eq:scalar}) for $\tilde f =
H_u,H_d$ with one of the Higgs fields replaced by its vacuum
expectation value.

The couplings to the gauge bosons are
\begin{eqnarray}
{\cal L}_{\rm int}^{(A)}
= {1 \over \sqrt 2}{ 2 g^2_A N \over (4 \pi)^2 }{1
\over \langle s
\rangle } 
\cdot {1 \over 4}s_R F^{\mu \nu}_{(A)} F_{(A) \mu \nu},
\end{eqnarray}
where the index $A$ represents the gauge group (SU(3), SU(2), and U(1)).

The main decay mode of $s_R$ depends on the mass spectrum of SUSY
particles. We discuss the case of the sweet spot SUSY
model~\cite{Ibe:2007km} in detail as an example of realistic models.
In this model, the $\mu$-parameter and the Higgs soft masses are
generated at the GUT scale through direct couplings between the Higgs
fields and the SUSY breaking sector. Those contributions do not depend
on $\langle s \rangle$. Additional large contributions to the $H_u$ soft
mass are generated through gauge mediation and the renormalization group
(RG) running. In particular, there is a significant RG effect due to the
large Yukawa coupling of the top quark and the large scalar top
masses. These contributions are proportional to $1/|\langle s
\rangle|^2$, and thus enhance the effective coupling to $s_R$.
The effective mass parameter $(m_{\rm eff}^{(H_u)})^2$ is estimated to be
\begin{eqnarray}
- ( m_{\rm eff}^{(H_u)} )^2 = (\kappa m_{\tilde B} )^2,
\end{eqnarray}
with
\begin{eqnarray}
\kappa \simeq 3 - 4.
\label{eq:kappa}
\end{eqnarray}
The parameter $\kappa$ depends logarithmically on the messenger scale.
For the down type Higgs,
\begin{eqnarray}
| (m_{\rm eff}^{(H_d)})^2| \ll  |(m_{\rm eff}^{(H_u)})^2|,
\end{eqnarray}
due to relatively small RG effects.  The effective coupling to the
Higgsinos is also suppressed,
\begin{eqnarray}
|\mu_{\rm eff}| \ll |\mu|,
\end{eqnarray}
in this model.

The enhancement in Eq.~(\ref{eq:kappa}) is very important since there is
no such factor in the $s_I$ decay. The decay modes $s_R \to hh$, $ZZ$,
and $WW$ (where the gauge bosons are longitudinally polarized) and also
the fermion modes such as $s_R \to b \bar b$ through the $s_R$-$h^0$
mixing are enhanced.
This makes the $s_R$ decay much faster than that of $s_I$ in the
parameter region of our interest.

The partial decay width of the $hh+WW+ZZ$ mode is given by
\begin{eqnarray}
\Gamma_{s_R \to hh} 
+ \Gamma_{s_R \to ZZ} 
+ \Gamma_{s_R \to WW} 
\simeq
{1 \over 8 \pi m_S}
\left(
{ \sqrt 2 (m_{\rm eff}^{(H_u)})^2  \sin^2 \beta
\over \langle s \rangle }
\right)^2.
\end{eqnarray}
We ignored the mass differences among $h$, $Z$ and $W$, and also ${\cal
O}(m_h^2 / m_S^2)$ and ${\cal O}(m_h^2 / m_A^2)$ terms for
simplicity. The angle $\beta$ is defined by $\tan \beta = \langle H_u
\rangle / \langle H_d \rangle$, and $m_A$ is the mass of the
pseudo-scalar Higgs boson.  If $s_R$ mainly decays into $hh+WW+ZZ$, the
decay temperature $T_d^{(R)}$ is given by
\bea
T_d^{(R)}&\simeq& 13{\rm \,\,GeV}\, \cdot N^{-1} 
\lrfp{g_*}{15}{-1/4} \lrfp{m_{\rm eff}^{(H_u)}}{800{\rm \,GeV}}{2}
\lrfp{m_S}{500 {\rm\,GeV}}{-1/2} \lrfp{m_{3/2}}{30{\rm\,MeV}}{-1} \non\\
&&\times \lrf{m_{\tilde B}}{200{\rm\,GeV}}\,  \sin^2 \beta ,
\eea
where we have defined the decay temperature as\footnote{Note that this
is a temperature of the Universe when the age of the Universe is
comparable to the lifetime of $s_R$, provided that the Universe is
radiation dominated or radiation produced by the $s_R$ decay dominates
over the Universe. Although $T_d^{(R)}$ does not represent a temperature
if the decays happen during the $s_I$ dominated era, we call this
quantity the decay temperature in later discussion even in that case.
}
\beq
T_d^{(R)} \;\equiv\; \lrfp{\pi^2 g_* }{ 90}{-1/4} \sqrt{\Gamma_{s_R} M_{\rm
Pl}}.
\label{eq:tdr}
\eeq
In the numerical calculations below, we have included the phase space
factors and contributions from other decay modes such as $t \bar t$.

The width of the $b \bar b$ mode through the mixing to the lightest
Higgs boson is given by
\begin{eqnarray}
\Gamma_{s_R \to b \bar b} \simeq 
{3 m_S \over 8 \pi}
\left(
{\sqrt 2 (m_{\rm eff}^{(H_u)})^2 \sin^2 \beta
\over
\langle s \rangle
}
\cdot 
{m_b \over m_h^2 - m_S^2}
\right)^2
%\left(
%1 - {4 m_b^2 \over m_S^2}
%\right)^{3/2}
.
\label{eq:widthbb}
\end{eqnarray}
Correction terms of ${\cal O}(m_b^2 / m_S^2)$ are ignored.
When the $b \bar b$ mode is dominant, the decay temperature is given by
\bea
T_d^{(R)}&\simeq& 1.6{\rm \,\,GeV}\, \cdot N^{-1} 
\lrfp{g_*}{15}{-1/4} \lrfp{m_{\rm eff}^{(H_u)}}{800{\rm \,GeV}}{2}
\lrfp{m_h}{115{\rm\,GeV}}{-2}
\lrfp{m_S}{100 {\rm\,GeV}}{1/2}  \non\\
&&\times \lrfp{m_{3/2}}{30{\rm\,MeV}}{-1} \lrf{m_{\tilde B}}{200{\rm\,GeV}}
\left| 1-\frac{m_S^2}{m_h^2} \right|^{-1} 
%\left(1-\frac{4 m_b^2}{m_S^2}\right)^{3/4} 
\,  \sin^2 \beta.
\eea
\begin{figure}[t]
\begin{center}
\includegraphics[height=7cm]{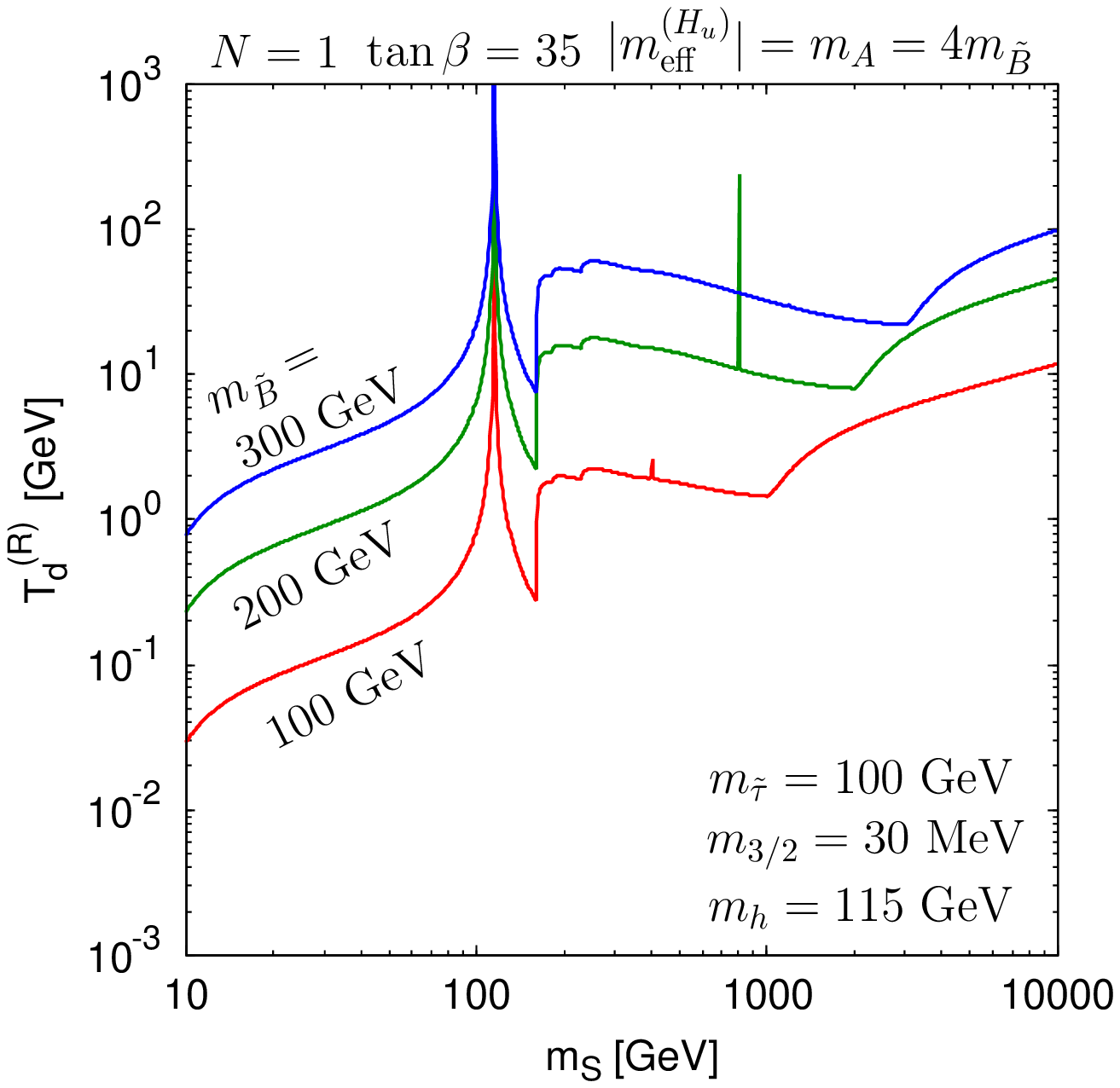}
\includegraphics[height=7cm]{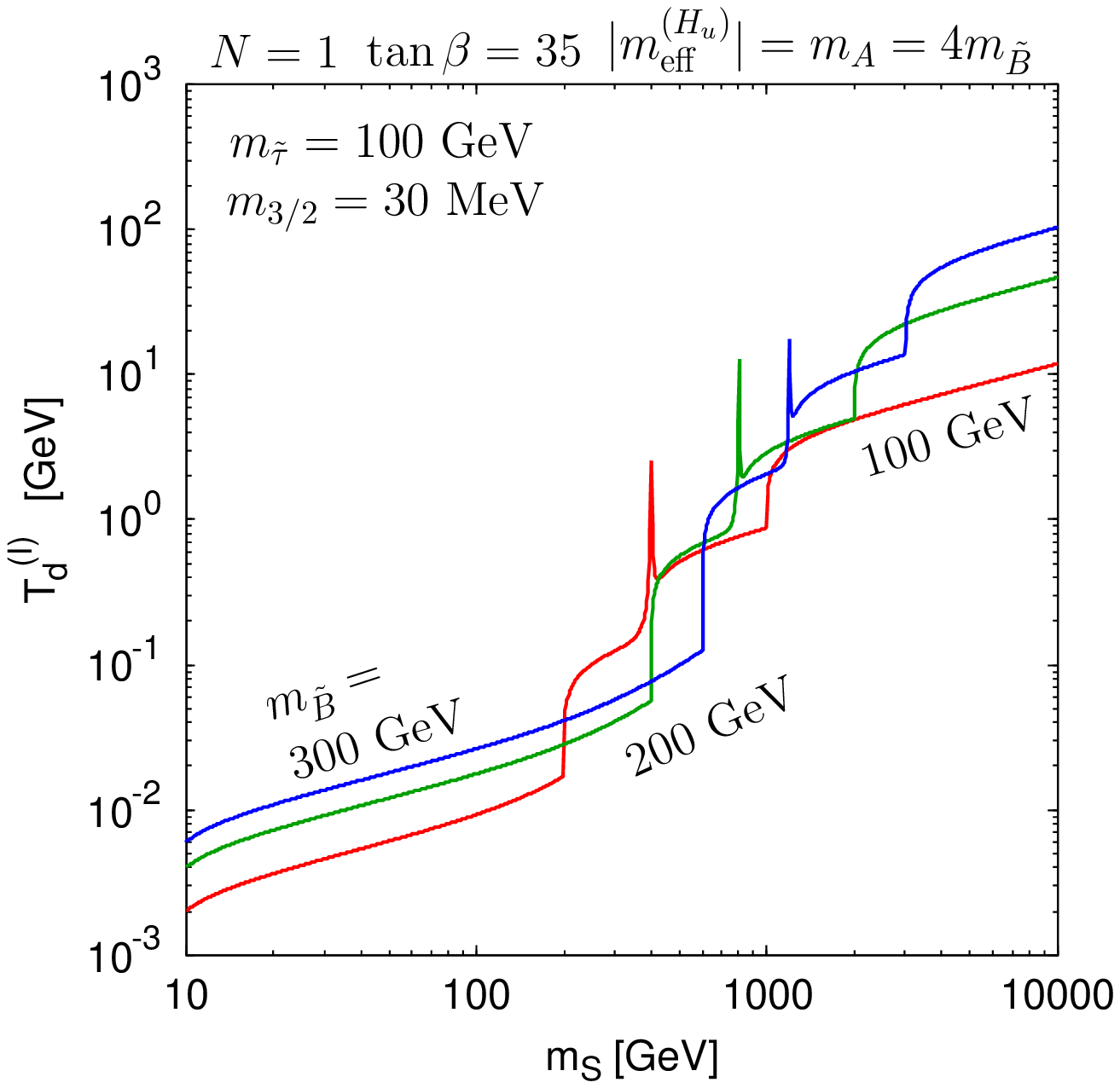}
\end{center}
\caption{The decay temperatures of $s_R$ and $s_I$. We fixed the
gravitino mass to be 30~MeV.  For other values of $m_{3/2}$, the 
decay temperatures are obtained by multiplying $(m_{3/2}/30~{\rm
MeV})^{-1}$.}  
\label{fig:temp}
\end{figure}

We show in the left panel of Fig.~\ref{fig:temp} the decay temperature
$T_d^{(R)}$ as a function of $m_S$ for $m_{\tilde B} = 100$, 200,
300~GeV.  In the figure, we fixed the gravitino mass to be 30~MeV. The
decay temperatures for other values of $m_{3/2}$ can be obtained by
multiplying a factor of $(m_{3/2}/30~{\rm MeV})^{-1}$.  We have fixed
other parameters to be $N=1$, $\tan \beta = 35$, $m_h = 115$~GeV and
$|m_{\rm eff}^{(H_u)}| = m_A = 4 m_{\tilde B}$, where $m_A$ is the
pseudo-scalar Higgs boson mass.  In the calculation, we have included
the decay modes $s_R \to hh, ZZ, WW, \gamma \gamma, \widetilde B
\widetilde B, \widetilde W \widetilde W, \tilde g \tilde g, t\bar t, b
\bar b, \tau \tau, \tilde \tau \tilde \tau, gg$.
The main decay mode of $s_R$ is $s_R \to hh,WW,ZZ$ for $2 m_W < m_S
\lesssim 1$~TeV. The gaugino modes become important for $m_S \gtrsim
1$~TeV. For $m_S < 2 m_W$, the $s_R \to b \bar b$ decay through the
$s_R$-$h$ mixing is the main decay process. We can see a sharp peak at
$m_S = m_h$ due to the enhancement of the mixing.
%

%%%%%%%%%%%%%%%%%%%%%%%%%%%%%
\subsection{Decays of {\boldmath $s_I$}}
%%%%%%%%%%%%%%%%%%%%%%%%%%%%%
In Ref.~\cite{Ibe:2006rc}, the decay property of $s_I$ is assumed to
be the same as the one of $s_R$. We show in this subsection that the
decay of $s_I$ happens much later in particular when $s_R \to hh$ is
open. The difference of the decay temperatures will be important in
calculating the non-thermal gravitino abundance.

The imaginary part, $s_I$, can only couple to CP-odd combinations. The
enhanced coupling to Higgs bosons through a large value of $m_{\rm
  eff}^{(H_u)}$ is therefore absent.
There is a coupling to the Higgs bosons through an $\langle s \rangle$
dependence of the $B\mu$-term:
\begin{eqnarray}
{\cal L} \ni B \mu H_u H_d + {\rm h.c.}, 
\to
{i \over 2 \sqrt 2} {m_A^2 \sin 2 \beta \over \langle s \rangle} s_I
H_u H_d + {\rm h.c.}
\label{eq:bmu}
\end{eqnarray}
Here we have assumed $B \mu \propto \langle s \rangle^{-1}$ and used a
tree-level relation from electroweak symmetry breaking, $B \mu = -
(m_A^2 \sin 2 \beta) / 2$.
The coupling constant is suppressed by a $\sin 2 \beta$ factor which
is generically small in gauge-mediation models.

The decay into two gauginos is therefore important if it is
kinematically open. The interaction Lagrangian is given by
\begin{eqnarray}
{\cal L}_{\rm int} \ni { 1 \over \sqrt 2} 
{m_\lambda \over \langle s \rangle} \cdot
{1 \over 2}
s_I \bar{\lambda} i\gamma_5 \lambda.
\end{eqnarray}
Since this is the same strength as the $s_R \bar \lambda \lambda$
coupling, the lifetime of $s_I$ is always longer than $s_R$ due to the
suppression of the Higgs modes.
The partial decay width of the Bino mode is given by
\begin{eqnarray}
\Gamma_{s_I \to \tilde B \tilde B} =
{m_S \over 32 \pi}
\left(
m_{\tilde B} \over \langle s \rangle
\right)^2
\left(
1 - {4 m_{\tilde B}^2 \over m_S^2}
\right)^{1/2}.
\end{eqnarray}
If this is the dominant decay channel, the decay temperature is given by
\begin{eqnarray}
T_d^{(I)} 
&\simeq& 720~{\rm MeV} 
\cdot
N^{-1}
\left( g_* \over 15 \right)^{-1/4}
\left( m_S \over 500~{\rm GeV} \right)^{1/2}
\left( m_{3/2} \over 30~{\rm MeV} \right)^{-1}
\left( m_{\tilde B} \over 200~{\rm GeV} \right)^{2}
\nonumber \\
&&
\times 
\left(
1 - {4 m_{\tilde B}^2 \over m_S^2}
\right)^{1/4},
\label{eq:tI}
\end{eqnarray}
where the decay temperature is defined in the same way as
(\ref{eq:tdr}).  The Binos subsequently decay into staus if $m_{\tilde
  B}>m_{\tilde \tau}$.  In a large $m_S$ region, the Wino and the
gluino modes become more important. However, as we will see later,
such a region is not allowed because of the overproduction of
gravitinos.

If the Bino mode is closed, the main decay mode is into $b \bar b$
through the $s_I$-$A^0$ mixing from Eq.~(\ref{eq:bmu}). The partial
width is calculated to be
\begin{eqnarray}
\Gamma_{s_I \to b \bar b} = 
{3 m_S \over 16 \pi}
\left(
{m_A^2 \sin^2 \beta \over \langle s \rangle}
\cdot
{m_b \over m_A^2 - m_S^2 }
\right)^2
%\left(
%1 - {4 m_b^2 \over m_S^2}
%\right)^{1/2}
.
\label{si2bbbar}
\end{eqnarray}
When the $b \bar b$ mode is dominant, the decay temperature is
\begin{eqnarray}
T_d^{(I)} &\simeq& 16~{\rm MeV}
\cdot N^{-1} 
\left(
{g_* \over 15}
\right)^{-1/4}
\left(
m_S \over 100~{\rm GeV}
\right)^{1/2}
\left(
m_{3/2} \over 30~{\rm MeV}
\right)^{-1}
\left(
m_{\tilde B} \over 200~{\rm GeV}
\right) 
\nonumber \\
&&
\times 
\left|
1 - {m_S^2 \over m_A^2}
\right|^{-1}
\sin^2 \beta.
\label{eq:tIbb}
\end{eqnarray}
The value in front becomes 18~MeV once we include the $\tau \tau$ mode.

The decay temperature as a function of $m_S$ is shown in the right
panel of Fig.~\ref{fig:temp}. We have used a set of parameters which
are indicated in the figure. Again, for other values of $m_{3/2}$,
$T_d^{(I)} \propto m_{3/2}^{-1}$.
As we can see, $T_d^{(I)}$ is significantly lower than $T_d^{(R)}$ for
$m_S \lesssim 1$~TeV.\footnote{Two decay temperatures become similar
when the $gg$ mode becomes dominant, i.e., $m_S < 2 m_\tau$.} The decay
modes $s_I \to \widetilde B \widetilde B, \widetilde W \widetilde W,
\tilde g \tilde g, t\bar t, b \bar b, \tau \tau, gg, \gamma \gamma, hA,
HZ$ are included in the calculation.  We have ignored the $hZ$ mode
because it is much smaller than the $b {\bar b}$ mode. The gauge
invariance requires the $hZ$ mode to vanish in the decoupling limit,
$m_h/m_A \to 0$.

\section{Non-thermal gravitino production}
\label{sec:gravitino}

The $s$ field can decay into two gravitinos with a suppressed
branching fraction. We calculate here the branching ratio and estimate
the gravitino energy density. We will see that the non-thermal
component can explain the DM abundance when $m_S \sim {\cal
  O}(100)$~GeV independent of the gravitino mass.

\subsection{Abundance}
The non-thermal gravitino abundance can be calculated from the decay
temperatures and the branching ratios of the $s_{R,I} \to \psi_{3/2}
\psi_{3/2}$ decays.  The partial decay width of $s_{R,I}$ into two
gravitinos, $s_{R,I} \to \psi_{3/2} \psi_{3/2}$, is given
by~\cite{Endo:2006zj,Nakamura:2006uc,Dine:2006ii,Ibe:2006rc}
\begin{eqnarray}
\Gamma_{3/2} = {1 \over 96 \pi} {m_S^3 \over M_{\rm Pl}^2 }
\left(
{m_S \over m_{3/2}}
\right)^2.
\end{eqnarray}
This formula is obtained from the interaction term in the second term
of Eq.~(\ref{eq:kahler}) by identifying the fermion component of $S$
with the longitudinal mode of the gravitino.
By using this partial decay width, we can calculate the branching
fraction.

There are two interesting branches where the main decay modes are
different. For the $s_R$ decay, the main decay mode is (A) $s_R \to b
\bar b$ for $2 m_b < m_S < 2 m_W$, and (B) $s_R \to hh, WW, ZZ$ for $2
m_W < m_S \lesssim 1$~TeV.
The branching ratios of the two-gravitino mode in those cases are
respectively given by
\begin{eqnarray}
B_{3/2}^{(R)} &\simeq&
4.6 \times 10^{-9} \cdot N^{2}
\left(|m_{\rm eff}^{(H_u)} | \over 800~{\rm GeV} \right)^{-4}
\left(m_S \over 100~{\rm GeV} \right)^{4}
\left(m_{\tilde B} \over 200~{\rm GeV} \right)^{-2}
\lrfp{m_h}{115 {\rm GeV}}{4}
\nonumber \\
&& \times
\left(1-\frac{m_S^2}{m_h^2}\right)^2
\sin^{-4} \beta
\ \ \cdots\mbox{(A)},
\nonumber 
\label{eq:brR}
\end{eqnarray}
\begin{eqnarray}
B_{3/2}^{(R)} \simeq
2.2  
\times 10^{-7}
\cdot
N^{2}
\left( | m_{\rm eff}^{(H_u)} | \over 800~{\rm GeV} \right)^{-4} 
\left( m_S \over 500~{\rm GeV} \right)^{6}
\left( m_{\tilde B} \over 200~{\rm GeV} \right)^{-2}
\sin^{-4} \beta
\ \cdots\mbox{(B)}.
\nonumber 
\label{eq:brR2}
\end{eqnarray}
It is interesting to notice that the branching ratio is independent of
$m_{3/2}$. 
For the $s_I$ decay, the main decay mode is 
(C) $s_I \to b \bar b$ ($m_S < 2 m_{\tilde B}$) or
(D) $s_I \to \widetilde B \widetilde B$ ($m_S > 2 m_{\tilde B}$). The
branching ratios in two cases are
\begin{eqnarray}
B_{3/2}^{(I)} \simeq
4.2 
\times 10^{-5}
\cdot
N^{2}
\left( m_S \over 100~{\rm GeV} \right)^{4}
\left( m_{\tilde B} \over 200~{\rm GeV} \right)^{-2}
\left(
1 - {m_S^2 \over m_A^2}
\right)^2
\sin^{-4} \beta
\ \ \cdots\mbox{(C)},
\nonumber 
\label{eq:brI}
\end{eqnarray}
\begin{eqnarray}
B_{3/2}^{(I)} \simeq
7.2
\times 10^{-5} \cdot 
N^2
\left( m_S \over 500~{\rm GeV} \right)^{4}
\left( m_{\tilde B} \over 200~{\rm GeV} \right)^{-4}
\left(
1 - {4 m_{\tilde B}^2 \over m_S^2}
\right)^{-1/2}
\ \ \cdots\mbox{(D)}
.
\nonumber
\label{eq:brI2}
\end{eqnarray}

We here define quantities $\Omega_{3/2}^{(R)}$ and $\Omega_{3/2}^{(I)}$
which represent the density parameters of the gravitino when we ignore
the presence of $s_I$ and $s_R$, respectively:
\begin{eqnarray}
\Omega_{3/2}^{(R)} &\equiv&
{3 \over 4} {m_{3/2}}
{T_d^{(R)} \over m_S} B_{3/2}^{(R)} \times 2
\Big/ (\rho_c/s)_0,
\end{eqnarray}
\begin{eqnarray}
\Omega_{3/2}^{(I)} &\equiv&
{3 \over 4} {m_{3/2}}
{T_d^{(I)} \over m_S} B_{3/2}^{(I)} \times 2
\Big/(\rho_c/s)_0,
\end{eqnarray}
where $(\rho_c/s)_0 \simeq 1.8 \times 10^{-9}$\,GeV is the critical
density divided by the entropy density at present.  The abundances
$\Omega_{3/2}^{(R)}$ and $\Omega_{3/2}^{(I)}$ are related as
\begin{eqnarray}
{ \Omega_{3/2}^{(R)} \over \Omega_{3/2}^{(I)} } = 
{ T_d^{(I)} \over T_d^{(R)}},
\label{eq:ratio}
\end{eqnarray}
where we used the fact that $B_{3/2}^{(R)} / B_{3/2}^{(I)} = (T_d^{(I)} /
T_d^{(R)})^2$.
In the actual situation, of course, one cannot totally neglect  $s_R$ or $s_I$, and
one has to
% take into account  both of the scalar fields, in general.
take into account both of the contributions and also the dilution effects.
The gravitino abundance in a general case can be expressed in terms of  
$\Omega_{3/2}^{(R)}$ and $\Omega_{3/2}^{(I)}$
as
\begin{eqnarray}
\Omega_{3/2}^{\rm NT}
= \left \{
\begin{array}{ll}
\Omega_{3/2}^{(R)}  
+
r^2 \Omega_{3/2}^{(I)}
\left( { \displaystyle{ T_d^{(R)} \over T_d^{(I)} }} \right),
& 
%\left(
{\rm for~~} r < \displaystyle{ \sqrt{ T_d^{(I)} / T_d^{(R)} } }
%\right)
\\
r^{-2}  \Omega_{3/2}^{(R)} 
\left( { \displaystyle{ T_d^{(I)} \over T_d^{(R)} }} \right)
+ \Omega_{3/2}^{(I)},
&
% \left(
{\rm for~~} r > \displaystyle{ \sqrt{ T_d^{(I)} / T_d^{(R)} } }
%\right)
\end{array}
\right..
\label{eq:omega}
\end{eqnarray}
The former and latter regions of $r$ respectively correspond to the
cases where $s_I$ does not and does dominate the energy density of the
Universe before the $s_I$ decays.  For $r > T_d^{(I)} / T_d^{(R)}$,
most of the gravitinos are produced by the $s_I$ decay. For $r^2 >
T_d^{(I)} / T_d^{(R)}$, both radiation and gravitinos arise from the
$s_I$ decay, and thus the gravitino abundance becomes insensitive to
$r$. Note that $\Omega_{3/2}^{(R)}$ ($\Omega_{3/2}^{(I)}$) corresponds
to the gravitino density parameter in the limit of $r \rightarrow 0$
($r \rightarrow \infty$).

Since $T_d^{(R,I)} \propto 1/\langle s \rangle \propto m_{3/2}^{-1}$ and
the branching fractions $B_{3/2}^{(R,I)}$ are independent of $m_{3/2}$,
both $\Omega_{3/2}^{(R)}$ and $\Omega_{3/2}^{(I)}$ are also independent
of the gravitino mass. Therefore, interestingly, the total gravitino
energy density $\Omega_{3/2}^{\rm NT}$ does not depend on the gravitino
mass.

%%%%%%%
\begin{figure}[t]
\begin{center}
\includegraphics[height=7cm]{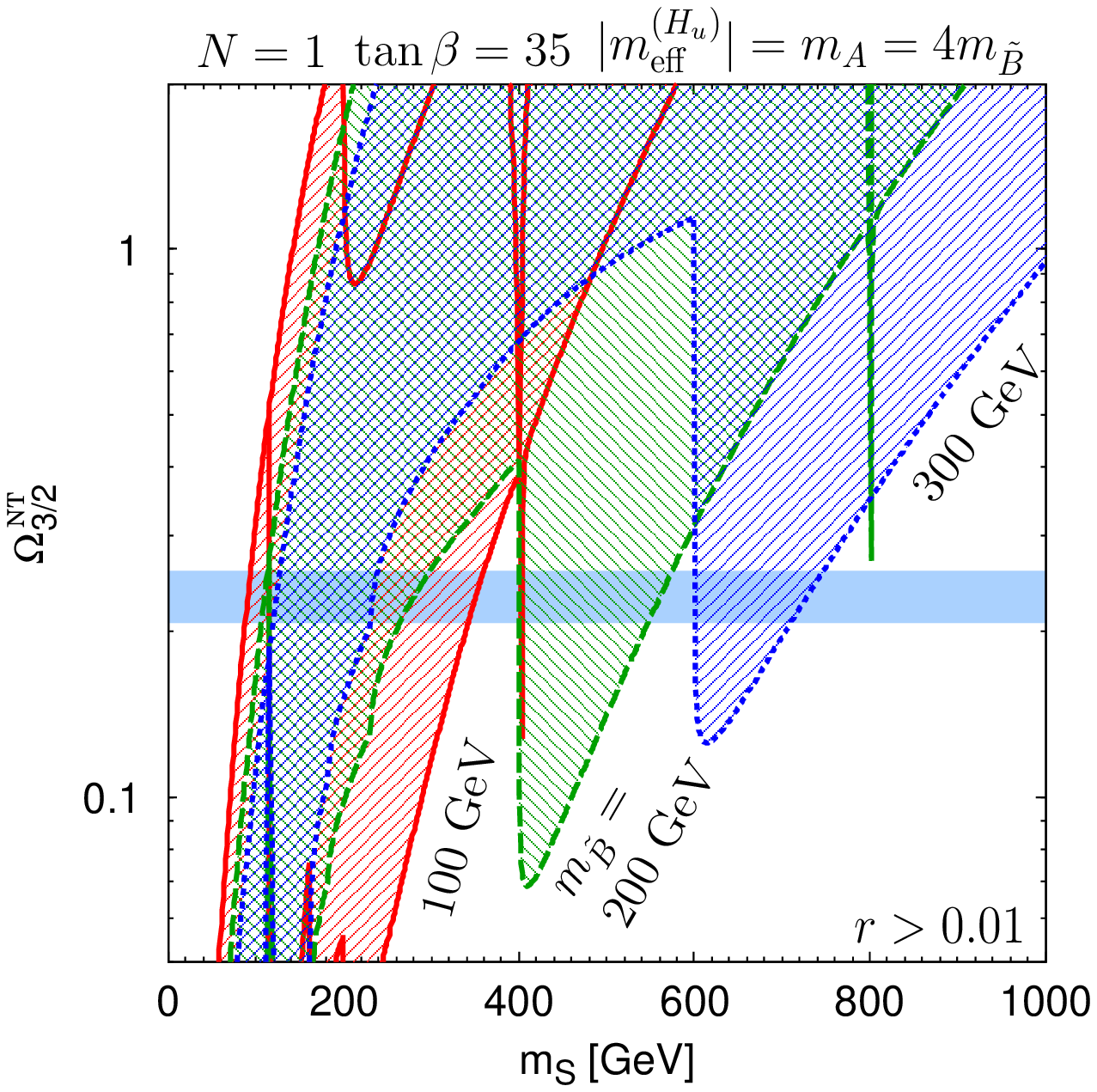}
\includegraphics[height=7cm]{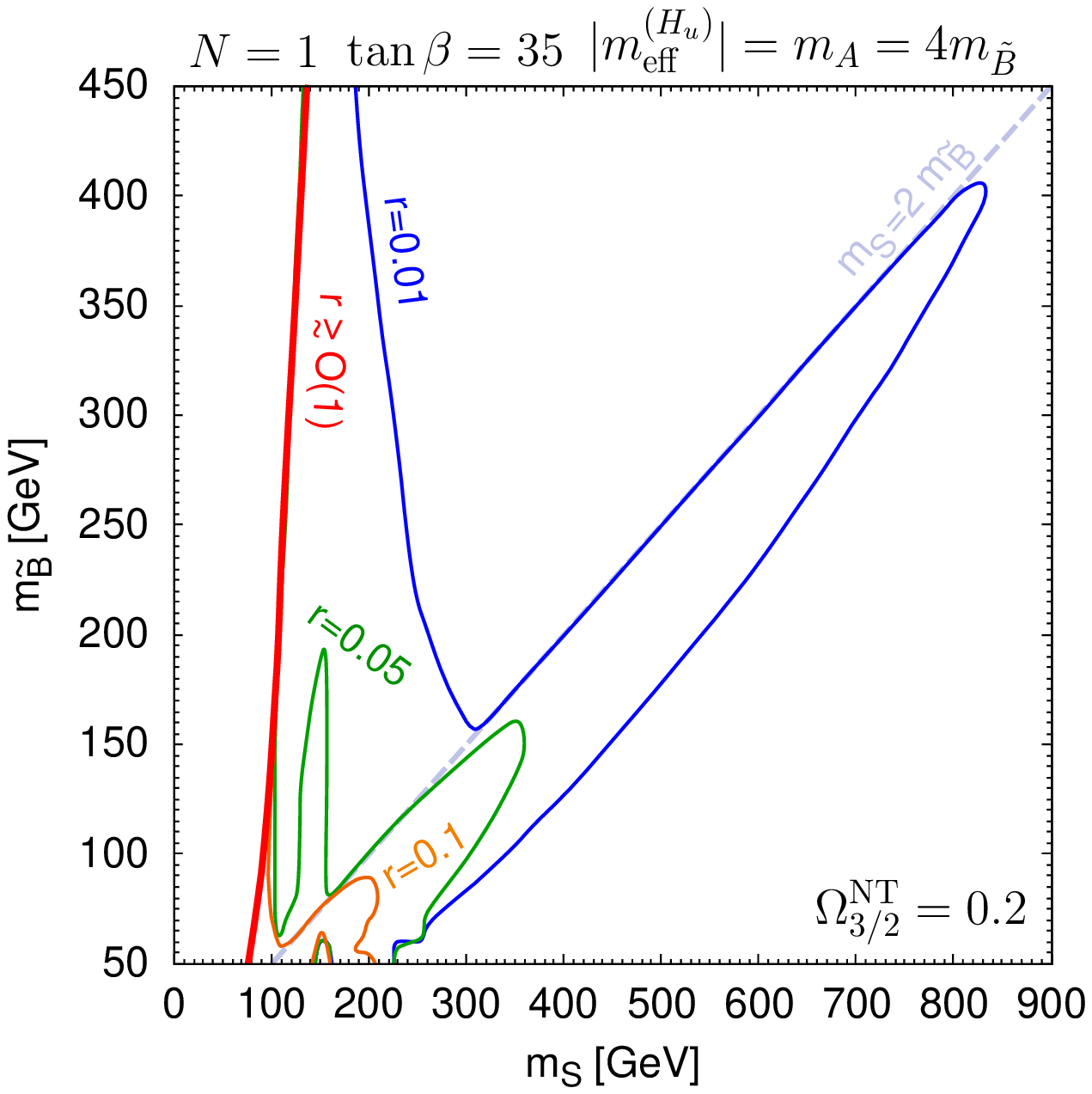}
\end{center}
\caption{The non-thermal gravitino abundance with respect to $m_S$ for
  several values of the Bino mass $m_{\tilde B} = 100, 200$, and
  $300$\,GeV (Left).  We have varied $r$ from $0.01$ to ${\cal O}(1)$
  for each value of $m_{\tilde B}$.  The other parameters are shown
  above the panel. For $r \gtrsim 1$, the three regions overlap one
  another around $m_S \sim 100$\,GeV and $m_{\tilde B} >
  100$\,GeV. The right panel shows the contours of $\Omega_{3/2}^{\rm
    NT} = 0.2$ for various values of $r$. For $r \gtrsim {\cal
    O}(0.1)$, the abundance does not depend on $r$ since the $s_I$
  oscillation dominates over the Universe. The gravitino abundance
  becomes larger as $m_S$ increases (see Eq.~(\ref{eq:omegaNT})). }
\label{fig:omega}
\end{figure}
%%%%%%

We show in the left panel of Fig.~\ref{fig:omega} the gravitino
abundance for fixed values of $m_{\tilde B}$ for $r>0.01$. The
non-thermal gravitino can account for the observed DM abundance for
$m_S = {\cal O}(100)$~GeV.
In the right panel, contours of $\Omega_{3/2}^{\rm NT}=0.2$ for
various values of $r$ are shown on the $m_S$-$m_{\tilde B}$ plane. For
$r \gtrsim {\cal O}(0.1)$, the contour gets independent of $r$ for
$m_{\tilde B} > 100$\,GeV. For $r \gtrsim 0.05$, the correct abundance
is obtained for $m_S \sim 100$~GeV where $s_I \to b \bar b$ is the
dominant decay mode. The gravitino abundance for that case is given by
\begin{eqnarray}
\Omega^{\rm NT}_{3/2} \simeq 0.2
\cdot
N
\left(
m_S \over 100~{\rm GeV}
\right)^{7/2}
\left(
m_{\tilde B} \over 200~{\rm GeV}
\right)^{-1}.
\label{eq:omegaNT}
\end{eqnarray}
The abundance does not depend on the detailed model parameters such as
$\kappa$, $m_A$ or $\tan \beta$ as long as $m_A \gg m_S$ and $\tan \beta
\gtrsim 3$.  In general, this result applies to models where the
$B\mu$-term is proportional to $\langle s \rangle^{-1}$ and the decay
width in Eq.~(\ref{eq:widthbb}) is larger than or comparable to that in
Eq.~\REF{si2bbbar}.

Although the abundance is independent of $m_{3/2}$, the gravitino mass
cannot be arbitrarily large. In order to avoid overproduction of
$^4$He, the decay temperature of $s_I$ is required to be higher than
$\sim 10$~MeV~\cite{Kawasaki:1999na}, which gives an upper bound on
the gravitino mass. For $m_S = 100$~GeV and $m_{\tilde B} \gsim
100$~GeV with $r \gtrsim 0.05$ motivated by DM abundance, we obtain
from Eq.~(\ref{eq:tIbb}):
\begin{eqnarray}
m_{3/2}\; \lesssim\; {\cal O}(100)~{\rm MeV}.
\label{eq:m32upper}
\end{eqnarray}

If one allows a small value of $r$, there is a region where $s_I \to
\tilde B \tilde B$ is open while $\Omega_{3/2}^{\rm total} = 0.2$ is
satisfied.
The upper bound on the gravitino mass in this case is relaxed to ${\cal
O}(1)$~GeV.
Such a region is subject to the BBN constraint from the decay of the
non-thermally produced NLSP. We will discuss the constraint later in
the Appendix~\ref{app:B}.

\subsection{Thermal component}
Here we comment on the amount of thermally produced gravitinos.
After accounting for the dilution effect by the entropy production of
the $s$ decays, the density parameter of the thermally produced
gravitinos is estimated to be~\cite{Ibe:2006rc}
\begin{eqnarray}
\Omega_{3/2}^{\rm th} &\simeq& 0.004
\left(
{m_{3/2} \over 30~{\rm MeV}}
\right)^{-1}
\left(
{m_{\tilde B} \over 200~{\rm GeV}}
\right)^2
\left(
{T_d \over 18~{\rm MeV}}
\right)
\left(
{|s_{\rm ini}| \over 2.5 \times 10^{15}~{\rm GeV}}
\right)^{-2},
\label{eq:thermal-grav}
\end{eqnarray}
where $|s_{\rm ini}|$ is the size of the initial amplitude. If the
entropy of the Universe is generated by the decay of $s_R$ ($s_I$),
one should substitute $T_d^{(R)}$ ($T_d^{(I)}$) for $T_d$.  This
expression is independent of the reheating temperature after inflation
even though most of the gravitinos are produced at the end of the
reheating process.\footnote{There is a logarithmic dependence on the 
reheating temperature through the running coupling.}

Let us assume that $s_I$ dominates the energy density of the Universe.
In the case where the $s_I \to b \bar b$ mode is the dominant decay
process, we obtain
\begin{eqnarray}
\Omega_{3/2}^{\rm th} &\simeq& 0.004 \cdot N^{-1}
\left(
{m_S \over 100~{\rm GeV}}
\right)^{5/2}
\left(
{m_{3/2} \over 30~{\rm MeV}}
\right)^{-4}
\left(
{m_{\tilde B} \over 200~{\rm GeV}}
\right)^3
\left(
{|s_{\rm ini}| \over \Lambda}
\right)^{-2},
\end{eqnarray}
where $\Lambda$ is the cut-off scale in Eq.~(\ref{eq:kahler}). The
$|s_{\rm ini}| / \Lambda$ factor cannot exceed ${\cal O}(1)$ for the
discussion to be within the framework of the effective theory.
In order for the thermal component not to exceed the observed DM
density of the Universe, we obtain a lower limit of the gravitino
mass:
\begin{eqnarray}
m_{3/2} \;\gtrsim\; {\cal O}(10)~{\rm MeV},
\label{eq:m32lower}
\end{eqnarray}
for $m_S \sim 100$~GeV  and $m_{\tilde B} \sim 200$\,GeV.

On the other hand, if $s_I \to {\tilde B} {\tilde B}$ mode is open,
the gravitino abundance becomes
\begin{eqnarray}
\Omega_{3/2}^{\rm th} &=& 0.1 \cdot N^{-1}
\left(
{m_S \over 500~{\rm GeV}}
\right)^{5/2}
\left(
{m_{3/2} \over 100~{\rm MeV}}
\right)^{-4}
\left(
{m_{\tilde B} \over 200~{\rm GeV}}
\right)^4 \non\\
&&\times
\left(
{|s_{\rm ini}| \over \Lambda}
\right)^{-2} \left(
1 - {4 m_{\tilde B}^2 \over m_S^2}
\right)^{1/4}.
\end{eqnarray}
The gravitino mass must be slightly heavier than the previous case, 
\begin{eqnarray}
m_{3/2} \gtrsim {\cal O}(100)~{\rm MeV},
\label{eq:m32lower2}
\end{eqnarray}
for $m_S \sim 500$~GeV and $m_{\tilde B} \sim 200$\,GeV.

\subsection{Free-streaming scale}

We have seen that the gravitinos non-thermally produced by the decay of
$s$ successfully account for the observed abundance of DM.
Since the gravitinos are relativistic at the production, we need to
check if the free-streaming scale is consistent with the observational
bound, $\lambda_{FS} \lsim {\cal O}(100)$\,kpc, from the Lyman
$\alpha$ forest data~\cite{Boyarsky:2008xj}.

Let us first derive an expression for the free-steaming length,
$\lambda_{FS}$, which is a distance that particles (gravitinos in our
case) can travel until they become non-relativistic. In the following
the gravitinos are assumed to be produced by the decay of $s$ (either
$s_R$ or $s_I$), and the decay temperature $T_d$ denotes either
$T_d^{(R)}$ or $T_d^{(I)}$.  Let us denote by $p$ the momentum of the
gravitino.  Due to the cosmic expansion, the momentum red-shifts as $p
\propto a^{-1}$, where $a$ denotes the scale factor.  Noting that the
velocity of a particle is given by the ratio of the momentum to the
energy, the free-streaming length is expressed as
\bea \lambda_{FS} &=& \int_{a_d}^{a_{eq}} \frac{p_d
  \lrf{a_d}{a}}{\sqrt{m_{3/2}^2 + p_d^2 \lrfp{a_d}{a}{2}} }\frac{1}{a}
\lrf{dt}{da} da,
\label{FS1}
\eea
where $p_d$ is the initial momentum at the production, $a_d$ and
$a_{eq}$ are the scale factors at the decay of $s$ and at the
matter-radiation equality, respectively, and the scale factor is
normalized to be unity at present. We take the matter-radiation
equality as the end point of integration, assuming that the gravitino
has become already non-relativistic at the equality. The assumption is
satisfied for the whole parameter space of interest.

Assuming that the Universe was radiation dominated since the gravitino
production until the equality, we can perform the integration and
obtain
\bea
\lambda_{FS}&\simeq& \frac{1+z_{eq}}{H_{eq}} \left(\frac{m_S}{2m_{3/2}} \frac{a_d}{a_{eq}} \right)
\sinh^{-1} \left(\frac{2m_{3/2}}{m_S} \frac{a_{eq}}{a_d} \right),\non\\
&\simeq& 60  {\rm\, kpc} \lrfp{g_*}{15}{-\frac{1}{4}} \lrfp{m_{3/2}}{30{\rm \,MeV}}{-1} 
\lrf{m_S}{100{\rm \,GeV}} \lrfp{T_d}{18{\rm\,MeV}}{-1} \non\\&&
\times \left\{1-0.1\ln\left[\lrfp{g_*}{15}{-\frac{1}{4}} \lrfp{m_{3/2}}{30{\rm \,MeV}}{-1}
\lrf{m_S}{100{\rm \,GeV}} \lrfp{T_d}{18{\rm\,MeV}}{-1}\right]\right\},
%1.8
%0.078
\label{FS2}
\eea
where $z_{eq}$ and $H_{eq}$ are the red-shift and the Hubble parameter
at the equality, respectively, and we have adopted an approximation,
$m_S \gg m_{3/2}$. In the last equality, we have used $z_{eq} \simeq
3176$ and $H_{eq} \simeq 31\, {\rm Mpc}^{-1}$~\cite{Komatsu:2008hk}.

If the decay of $s$ produces not only the gravitinos but also the (almost) entire
entropy of the Universe,
we can express the free-streaming length in terms of the
branching fraction of the gravitino production.  The free-streaming length is
then give by  
\bea
\lambda_{FS}&\simeq& 60 {\rm\, kpc} \lrfp{g_*}{15}{-\frac{1}{4}}
\lrfp{\Omega_{3/2}^{\rm NT}}{0.2}{-1}\lrf{B_{3/2}}{4\times 10^{-5}}\non\\
&&\times \left\{1-0.1\ln\left[ \lrfp{g_*}{15}{-\frac{1}{4}}
\lrfp{\Omega_{3/2}^{\rm NT}}{0.2}{-1}\lrf{B_{3/2}}{4 \times 10^{-5}}\right]\right\}.
\label{FS3}
%0.21
%0.079
\eea
Since $B_{3/2}$ is independent of $m_{3/2}$, the free-streaming scale is
also independent of $m_{3/2}$ once we fix the gravitino abundance.
Note that the expression \EQ{FS3} is valid only for $r <
T_d^{(I)}/T_d^{(R)}$ or $r > \sqrt{T_d^{(I)}/T_d^{(R)}}$, while \EQ{FS2}
holds for any values of $r$ as long as the gravitino comes mainly from
either $s_R$ or $s_I$.  
In the parameter region of our interest, where $s_I \to b \bar b$ is
the main decay mode, the free-streaming scale is of ${\cal
  O}(100)$~kpc, which is on the border of the Lyman-$\alpha$ bound,
$\lambda_{FS} \lsim {\cal O}(100)$\,kpc.  It is an interesting
possibility that we may be able to see the suppression of the
structure formation below the corresponding scale.

\subsection{Isocurvature perturbations}
During inflation, $s$ is assumed to be at $|s_{\rm ini}| \sim
\Lambda$, far deviated from the origin. If the $s$ field has an
approximate U(1) symmetry, the phase component, $\theta \equiv
{arg}[s]$, remains light and therefore acquires quantum fluctuations
$\delta \theta = H_I/(2\pi |s_{\rm ini}|)$, where $H_I$ represents the
Hubble parameter during inflation. The phase $\theta$ is related to
$r$ (the ratio of the initial values of $s_R$ and $s_I$) as $\tan
\theta = r$. Therefore, $\delta \theta$ amounts to the fluctuation
$\delta r$, which generically leads to the isocurvature fluctuations
in the gravitino DM. This is because $s_R$ and $s_I$ have different
decay temperatures and different branching ratios into the gravitinos.

Recall that the gravitino abundance becomes insensitive to $r$ for $r
\gg \sqrt{T_d^{(I)}/T_d^{(R)}}$.  This is because both the radiation
and the gravitino are produced mainly from the decay of $s_I$.  Thus,
the cold DM (CDM) isocurvature perturbation is also suppressed in this
case~\cite{Hamaguchi:2003dc}.  Intuitively speaking, for a large
enough value of $r$, we can simply neglect the $s_R$; the fluctuations
in radiation and the gravitino DM are then adiabatic, since both are
generated from a single source, $s_I$.

To see this more explicitly, let us estimate the CDM isocurvature
perturbation $S_{c \gamma}$ in the case of $r \gg
\sqrt{T_d^{(I)}/T_d^{(R)}}$.  From Eqs.~(\ref{eq:ratio}) and
(\ref{eq:omega}), we have
\beq
\ds{S_{c \gamma} \;\equiv\; \delta \left(\log\left(\frac{\rho_{3/2}}{s}\right) \right) \simeq
%\frac{-2  r^{-3} \lrfp{ T_d^{(I)}}{ T_d^{(R)}}{2}}{1+r^{-2}\lrfp{ T_d^{(I)}}{ T_d^{(R)}}{2}}}
%(1+r^2) \frac{H_I}{2 \pi |s_{\rm ini}},
-2  \frac{1+r^2}{r^{3}} \lrfp{ T_d^{(I)}}{ T_d^{(R)}}{2}} \frac{H_I}{2 \pi |s_{\rm ini}|},
\eeq
where we have used $\delta r = \delta\theta/\cos^2\theta$ in the last
equality. As we mentioned above, we can see that $S_{c \gamma}$ is
suppressed for $r \gg \sqrt{T_d^{(I)}/T_d^{(R)}}$.  The current
observation bound on the isocurvature perturbation reads $|S_{c
  \gamma}| \lesssim 2 \times 10^{-5}$ at
$95$\%C.L.~\cite{Komatsu:2008hk}.  Thus, $H_I$ is bounded above not to
exceed the current constraint on the isocurvature perturbation,
\beq
H_I \;\lesssim\;
%\frac{r^{3}}{1+r^2} \lrfp{ T_d^{(R)}}{ T_d^{(I)}}{2} \pi \Lambda \lrf{|s_{\rm ini}|}{\Lambda}2 \times 10^{-5}
2 \times \GEV{11} \lrf{r^{3}}{1+r^2} \lrfp{ T_d^{(R)}}{ T_d^{(I)}}{2} 
\lrf{m_{3/2}}{30~{\rm MeV}}
\lrfp{m_S}{100~{\rm GeV}}{-1}
 \lrf{|s_{\rm ini}|}{\Lambda}.
\eeq
There are plentiful inflation models which satisfy the bound.

\section{Sweet spot}

We have seen that the non-thermal production of the gravitino can
explain DM of the Universe in a class of gauge mediation models. For
$r \gtrsim 0.05$, the correct abundance is obtained when the $s_I \to
b \bar b$ mode is the dominant decay mode. The abundance in that case
is given by Eq.~(\ref{eq:omegaNT}) which is independent of $m_{3/2}$.
The constraints from the thermal production of the gravitino
(Eq.~(\ref{eq:m32lower})) and the decay temperature
(Eq.~(\ref{eq:m32upper})) restrict the mass range of the gravitino to be
\begin{eqnarray}
10~{\rm MeV} \lesssim m_{3/2} \lesssim 100~{\rm MeV}.
\end{eqnarray}
%Interestingly, this mass range overlaps with the prediction of the
%gravitational stabilization mechanism in Ref.~\cite{Kitano:2006wz}.
Interestingly, the above mass range and $m_S \sim 100~{\rm GeV}$
overlaps with  the prediction of the
gravitational stabilization mechanism in Ref.~\cite{Kitano:2006wz}.
This model relates $\langle s \rangle$ and $\Lambda$ by
\begin{eqnarray}
\langle s \rangle = {\sqrt 3 \Lambda^2 \over 6 M_{\rm Pl}},
\end{eqnarray}
which is translated into a relation among $m_S$, $m_{3/2}$ and
$m_{\tilde B}$ as
\begin{eqnarray}
m_{3/2} = 34~{\rm MeV} \cdot N
\left(
m_S \over 100~{\rm GeV}
\right)^2
\left(
m_{\tilde B} \over 200~{\rm GeV}
\right)^{-1}.
\end{eqnarray}
The reference value we took approximately satisfies the relation. It is
also interesting to note that the above supergravity effect always
exists. Therefore, there is no big room left for other mechanisms to
give messenger masses in the scenario of the $s$-dominated Universe.

Ref.~\cite{Ibe:2007km} proposed a solution to the $\mu$ problem by
using the above gravitational stabilization mechanism. The $\mu$-term
is generated from the direct interaction terms between the SUSY
breaking sector and the Higgs fields, $K \ni S^\dagger H_u H_d /
\Lambda$. This framework predicts
\begin{eqnarray}
\mu \sim {F_S \over \Lambda} \sim m_S,
\end{eqnarray}
which is perfectly consistent with $\mu \sim {\cal O}(100)$~GeV required from
electroweak symmetry breaking and $m_S \sim {\cal O}(100)$~GeV from gravitino
DM.

\section*{Acknowledgement}
We thank M.~Ibe for reading the manuscript and useful discussions.
The work of R.K.~is supported in part by the Grant-in-Aid for
Scientific Research (No. 18071001) from the Japan Ministry of
Education, Culture, Sports, Science and Technology. The work of
F.T. was supported by JSPS Grant-in-Aid for Young Scientists (B)
(21740160).  We would like to thank the IPMU focus week on LHC
physics, June 23-27, 2008, where this work has been initiated.  This
work was supported by World Premier International Center Initiative
(WPI Program), MEXT, Japan.

\appendix
\section{Spatial instabilities}
\label{app:si}
Let us estimate a condition for an instability to grow.
The scalar potential of $s = (s_R+i s_I)/\sqrt{2}$ is given by
\beq
V(s_R,s_I) \;= \; \frac{1}{2} m_S^2 \left( (s_R- \sqrt{2} \vev{s})^2 +s_I^2 \right)
 + \frac{5N}{16\pi^2}\lambda^2 F_S^2\log\lrf{s_R^2 + s_I^2 }{2 \Lambda^2}.
\eeq
Differentiating the scalar potential with respect to $s_R$ and $s_I$, we obtain
%%%
%\bea
%\frac{\partial V}{\partial s_R} &=& m_S^2 (s_R - \sqrt{2}\vev{s}) +  \frac{5N}{16\pi^2}\lambda^2 F_S^2 \frac{2 s_R}{s_R^2+s_I^2},\\
%\frac{\partial V}{\partial s_I} &=& m_S^2 s_I +  \frac{5N}{16\pi^2}\lambda^2 F_S^2 \frac{2 s_I}{s_R^2+s_I^2},
%\eea
%%%
%and
%%
\bea
{\cal M}^2 &\equiv&
\left(
\bear{cc}
\frac{\partial^2 V}{\partial s_R^2} & \frac{\partial^2 V}{\partial s_R \partial s_I} \\
\frac{\partial^2 V}{\partial s_R \partial s_I} & \frac{\partial^2 V}{\partial s_I^2}  \\
\eear
\right) \non\\
&=&{\ds
\left(
\bear{cc}
m_S^2 +  \frac{5N}{16\pi^2}\lambda^2 F_S^2 
\left( \frac{2 }{s_R^2+s_I^2} - \frac{4 s_R^2}{(s_R^2+s_I^2)^2} \right) &
 - \frac{5N}{16\pi^2}\lambda^2 F_S^2 \frac{4 s_R s_I}{(s_R^2+s_I^2)^2}\\
  - \frac{5N}{16\pi^2}\lambda^2 F_S^2 \frac{4 s_R s_I}{(s_R^2+s_I^2)^2}&
 m_S^2 +  \frac{5N}{16\pi^2}\lambda^2 F_S^2
\left( \frac{2 }{s_R^2+s_I^2} - \frac{4 s_I^2}{(s_R^2+s_I^2)^2} \right)
\eear
\right)
}\non\\
\eea
Neglecting the cosmic expansion, the instability grows if ${\rm det}[{\cal M}^2] < 0$.
Thus, the condition for the instabilities not to grow is ${\rm det}[{\cal M}^2] > 0$,
namely
\beq
|s|^2 > \frac{5N}{16\pi^2 m_S^2}\lambda^2 F_S^2 = \lrfp{\lambda}{ \lambda_{\rm max}}{2} \frac{\vev{s}^2}{4 },
\eeq
or equivalently,
\beq
\lambda \;<\; 2 \lambda_{\rm max} \sin \theta.
\eeq
%%

%%%%%%%%%%%%%%%%%%%%
\section{Remarks on initial conditions}
\label{app:T}
%%%%%%%%%%%%%%%%%%%%
In this Appendix, we discuss conditions for $s_{R,I}$ to dominate the
energy density of the Universe, taking account of finite temperature
effects. For concreteness, we set $m_{3/2}\simeq 30~{\rm MeV}$,
$m_{\tilde B}\simeq 200~{\rm GeV}$, $m_S\simeq 100~{\rm GeV}$ and $N=1$, and
also $r\gsim {\cal O}(1)$ as reference values, which lead to a
successful gravitino DM scenario from the $s_I$ decay, as discussed in
the text. Therefore we are concerned with a condition for $s_I$ to
dominate the energy density of the Universe.

%%%%%%%%%%%%%%%
\subsection{$s_I$-domination}
%%%%%%%%%%%%%%%
The $s_I$-domination condition \EQ{eq:sIdomination} leads to
\begin{eqnarray}
\left|{\rm Im}[s_{\rm ini}]\right| &\gsim & 
7\times \GEV{12}
\cdot
\lrfp{{\rm min}[T_R, T_{m_S}]}{7\times\GEV{9}}{-1/2}
\lrfp{T_d^{(I)}}{18\,{\rm MeV}}{1/2}.
\label{dom}
\end{eqnarray}
If we take a natural expectation, $|s_{\rm ini}|\sim \Lambda$, the $s_I$
domination can be realized for $T_R\gsim \GEV{5}$.

%%%%%%%%%%%%%%%%%%%%%%%%%%%%%%
\subsection{Finite temperature effects}
\label{app:finiteT}
%%%%%%%%%%%%%%%%%%%%%%%%%%%%%%

In this subsection,
we assume that the $s$ field starts its oscillations when the Universe
is dominated by the oscillating inflaton. Even before the reheating,
however, there is a background dilute plasma with a temperature
$T\simeq (T_R^2 M_{\rm Pl} H)^{1/4}$.  The potential of $s$ field
therefore receives thermal corrections, which are not taken into
consideration so far.  Here we briefly discuss the finite temperature
effects on the evolution of the $s$ field and the messenger fields.

There are two thermal effects on the $s$ field: thermal mass and
thermal logarithmic terms, which arise depending on whether the
messenger fields are in thermal bath or not.  If the effective masses
of the messenger fields are smaller than the temperature of thermal
plasma, i.e., $m_{\rm eff}=\lambda |s| < T$, the messenger fields will
be in thermal equilibrium.  The $s$ field then receives a thermal
mass:
\beq
V_T^{(1)} \;=\; 
\frac{5}{4}N \lambda^2 T^2 |S|^2 ~~~~~~~{\rm for~~}\lambda |s| < T,
\label{vt1}
\eeq
where we have assumed that the messenger fields transform as ${\bf 5}$
and $\bf{\bar 5}$ under SU(5).

On the other hand, when the messenger fields are so heavy that they
are decoupled from thermal bath, there is a thermal effect arising
from the two-loop contribution to the free energy, $\delta V\propto
g(T)^2 T^4$.  Here we consider only the SU(3)$_C$ gauge group, which
gives the dominant contribution to the free
energy~\cite{Buchmuller:2004xr},
\beq
\delta V \;=\; \frac{21}{8}g_3^2(T) T^4.
\eeq
For $\lambda |s|>T$, the running gauge coupling $g_3(T)$ is modified as 
\begin{eqnarray}
  g_3(T)|_{\lambda |s|>T} &=& g_3(T)|_{s=\vev{s}} + N \frac{g_3(M_U)^3}{32\pi^2} \ln \lrf{\lambda^2 |s|^2}{T^2},
\end{eqnarray}
where $M_U$ is some ultraviolet scale where $g_3$ is fixed.  This
leads to a thermal correction to the scalar potential
\begin{eqnarray}
V_T^{(2)}  &=& \frac{21N}{8}\alpha_3(T)^2 T^4 \ln\lrf{|s|^2}{T^2}
~~~~~~~{\rm for~~}\lambda |s| > T,
\label{vt2}
\end{eqnarray}
which may become important where the thermal mass term is negligible. 

Next let us consider the thermal effect on the messenger fields. The
messenger fields acquire thermal masses through the gauge interactions
with the SSM particles in thermal plasma. The thermal masses tend to
prevent the messengers from falling into a SUSY minimum. In principle
this effect could enlarge the allowed region for $r$: even with a
small value of $r$, the messengers may be stabilized at their origin
and the $s$ field may settle down at the SUSY breaking minimum in the
end. However, if this is the case, our scenario would be modified in
two ways. First, the messenger fields are in thermal equilibrium when
they are stabilized by their thermal masses. If the messenger number
is conserved, the lightest messenger may exceed the DM
abundance. Although this issue can be avoided by introducing the
breaking of the messenger number, it would make the analysis
model-dependent.  Second, the gravitino abundance
\REF{eq:thermal-grav} is modified because the gravitinos are also
generated from the scattering processes including
messengers~\cite{Jedamzik:2005ir}.  Thus, in order to keep the success
of our scenario in the text, we assume that the messenger fields are
so heavy that they are always decoupled from thermal plasma.

\begin{figure}[t!]
\begin{center}
\includegraphics[width=8cm]{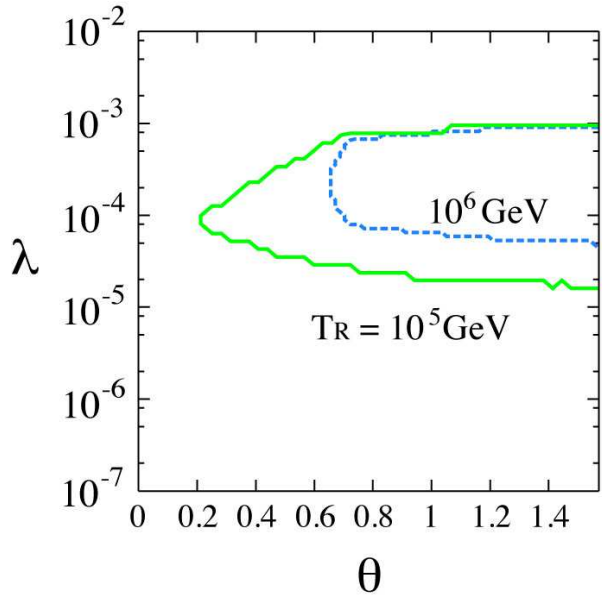}
\end{center}
\caption{The parameter region where the $s$ field is trapped by the
  SUSY breaking minimum and the messenger fields are decoupled from
  the thermal bath, when the thermal effects \REF{vt1} and \REF{vt2}
  are taken into account.  We have set $m_{3/2}=30~{\rm MeV}$,
  $m_{\tilde B}=200~{\rm GeV}$, $m_S=100~{\rm GeV}$ and $N=1$.}
\label{fig:allowed_w_T}
\end{figure}

We have shown in Fig.~\ref{fig:allowed_w_T} a parameter space in which
(i) the perturbation $\delta s$ remains small compared to ${\bar s}$
until the $s$ field is stabilized at the SUSY breaking minimum and (ii)
the messengers remain decoupled from thermal bath during the course of
evolution, for $T_R = \GEV{5}$ and $\GEV{6}$ with the thermal effects
\REF{vt1} and \REF{vt2} taken into account.  Here we have set
$m_{3/2}=30~{\rm MeV}$, $m_{\tilde B}=200~{\rm GeV}$, $m_S=100~{\rm
GeV}$ and $N=1$.  Notice that, compared to the zero-temperature result
shown in Fig.~\ref{fig:allowed}, smaller values of $\lambda$ are
excluded since the messengers would be thermalized. The allowed region
disappears for $T_R > \GEV{7}$.  Therefore, our scenario works for
$10^{-5} \lesssim \lambda \lesssim 10^{-3}$, $0.2 \lesssim \theta \leq
\pi/2$ and $\GEV{5} \lesssim T_R \lesssim \GEV{6}$, if thermal effects
are taken into account. Note that the consistent ranges for $\lambda$
and $\theta$ depend on the choice of $m_{3/2}$, $m_{\tilde B}$ and
$m_S$. For instance, the lowest allowed value of $\theta$ can be as
small as ${\cal O}(0.01)$ for e.g. $m_S = 400$\,GeV and $m_{\tilde B} =
100$\,GeV.

%%%figure%%%
%%%figure%%%

%%
\section{BBN constraints}
\label{app:B}

\begin{figure}[t]
\begin{center}
\includegraphics[height=7cm]{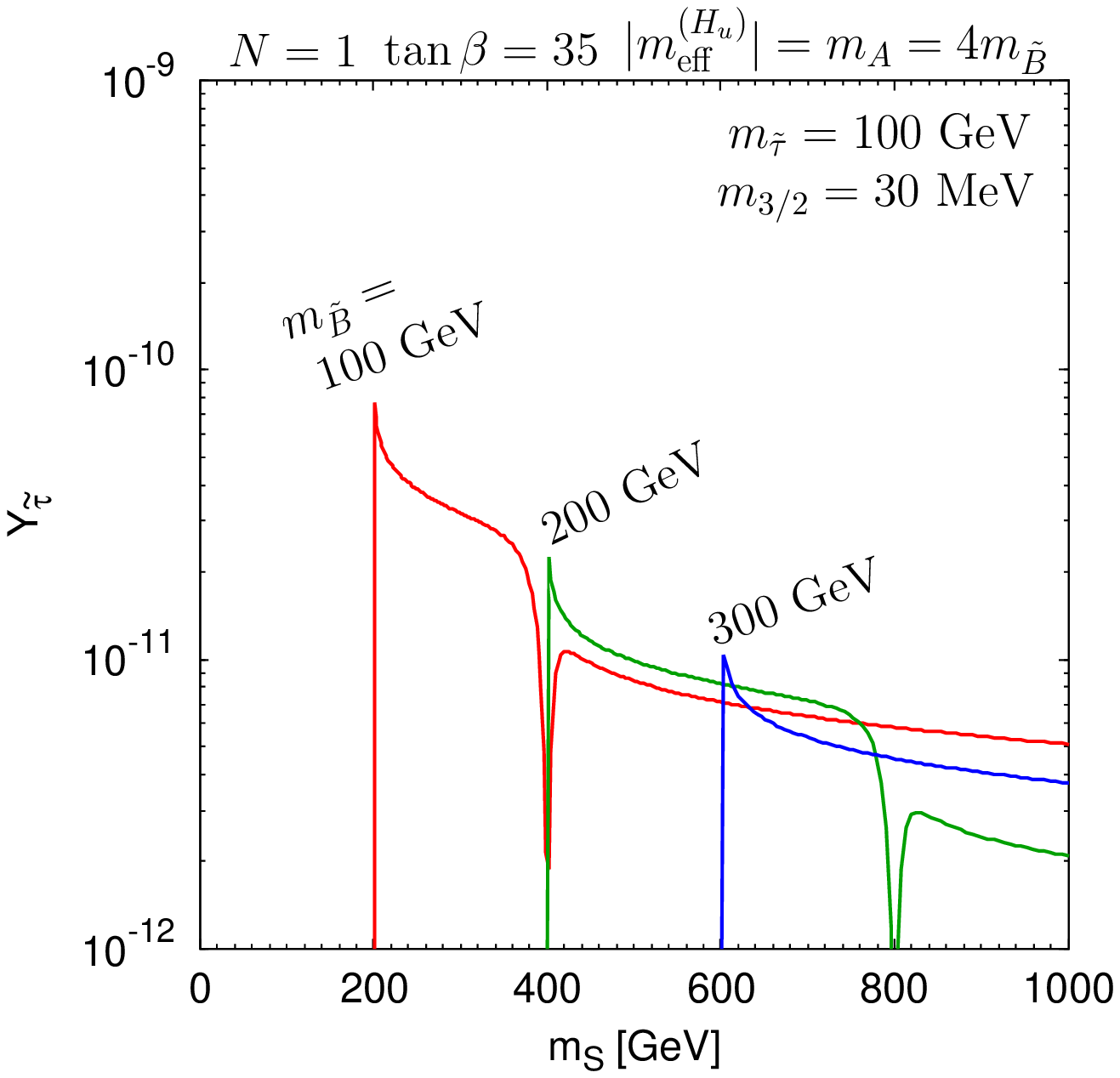}
\includegraphics[height=7cm]{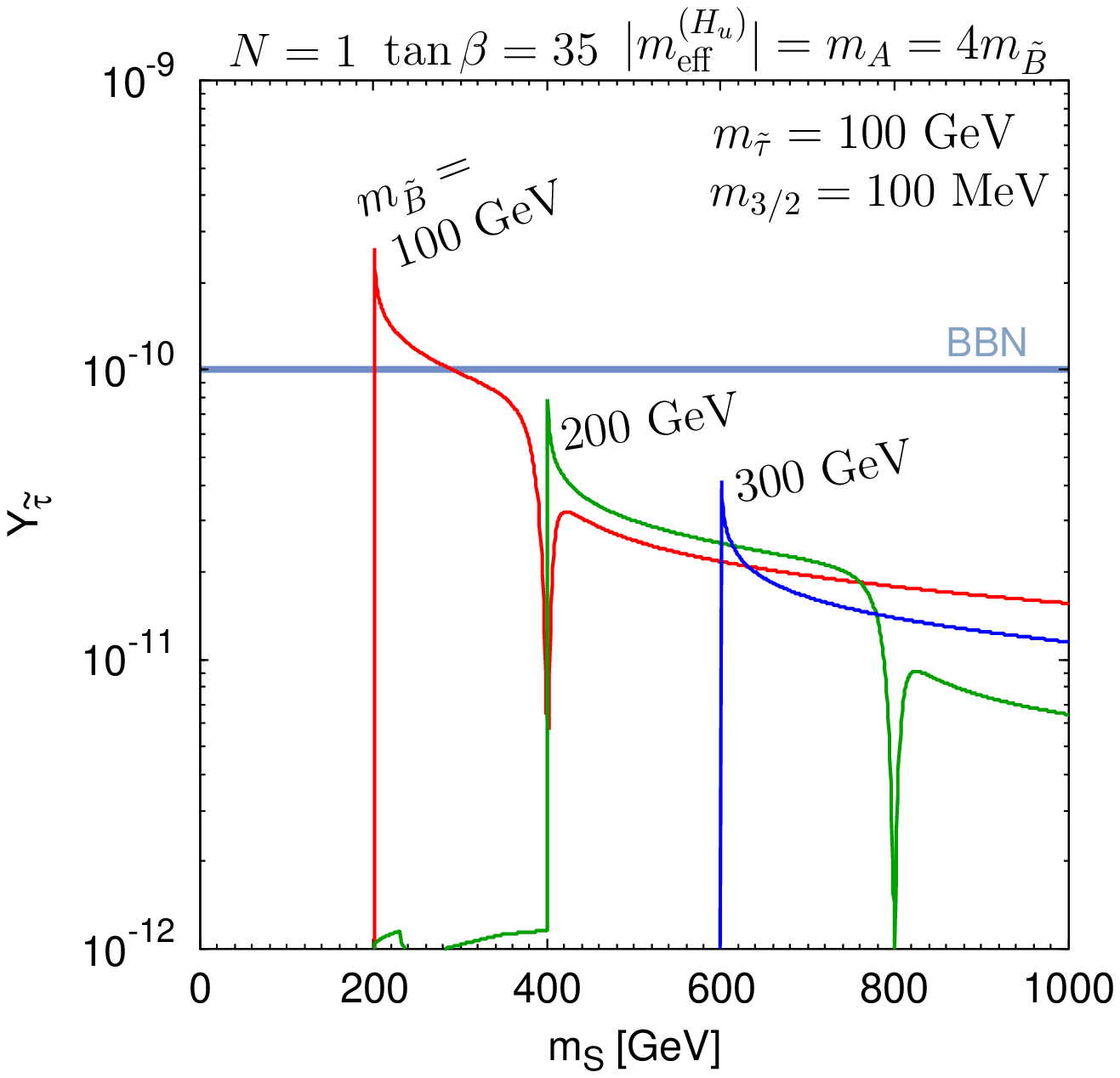}
\end{center}
\caption{The abundance of the non-thermally produced staus. We took
$m_{\tilde \tau}=100$~GeV. The gravitino mass is $m_{3/2}=30$~MeV
(left) and 100~MeV (right). The abundance is inversely proportional to
the decay temperature.}  \label{fig:Y-stau}
\end{figure}

If the decay of $s_R$ or $s_I$ into the superparticles are
kinematically allowed, they are copiously produced, which will decay
into the NLSP promptly. Depending on the lifetime of the NLSP, their
abundance is subject to the BBN constraint.  We assume in the
following that the NLSP is the stau since the constraint is much
weaker than the Bino NLSP case.

Before proceeding, let us mention when the BBN constraint could become
important. As one can see from the right panel in Fig.~\ref{fig:omega},
there are parameter regions where the decay into the superparticles is
significant for $r \ll 1$ while the gravitino abundance is fixed.  On
the other hand, for the reference values of $m_{3/2}=30~{\rm MeV}$,
$m_{\tilde B}=200~{\rm GeV}$, $m_S=100~{\rm GeV}$ and $N=1$, we have
found that $r$ must be larger than $0.2$ ($0.05$), if thermal effects
are (not) taken into consideration for our scenario to work (see
Figs.~\ref{fig:allowed} and \ref{fig:allowed_w_T}). However, for a
different choice of those parameters, the smallest value of $r$ can be
${\cal O}(0.01)$.  Thus, for a certain fraction of the parameter space
of our concern, the BBN constraint may be important.

The main source of stau is the $s_I \to \widetilde B \widetilde B$
decay followed by the decays of Binos into staus, if the Bino mode is
open. In this case, the decay of $s_R$ is not important since the
$s_I$ decays much later.
If the Bino mode is closed, the main source is $s_R \to \tilde
\tau \tilde \tau$.

By using the decay temperatures calculated before, the non-thermal
stau abundance can be estimated through (\ref{eq:NT}) in
Appendix~\ref{sec:NT}, where a general formula of the non-thermal
relic abundance is derived.  When a large number of staus are produced
by the $s_I$ decay, the fast pair annihilation processes make the
final abundance approach a value determined by the decay temperature
and the annihilation cross section, which is not sensitive to the
initial abundance.

We show in the left panel of Fig.~\ref{fig:Y-stau} the abundance of the
non-thermally produced stau as a function of $m_S$ with the same set of
parameters as Fig.~\ref{fig:temp}. The right figure is the case with
$m_{3/2}=100$~MeV. We have used the annihilation cross section of the
staus in Ref.~\cite{Asaka:2000zh}. (Recently it has been shown that the
cross section can be larger if there is a significant left-right mixing
in the stau sector~\cite{Pradler:2008qc,Ratz:2008qh}.)
The parameter $r$, the ratio of the amplitudes, is taken to be $r =
0.01$.
%, since a larger value of $r$ would result in the overproduction of
%the gravitinos for $m_S \gsim 200$~GeV as one can see in
%Fig.~\ref{fig:omega}.

The abundance $Y_{\tilde \tau}$ does not depend on $r$ if the $s_I \to
\widetilde B \widetilde B$ decay is kinematically allowed ($ m_S > 2
m_{\tilde B}$). In the case where $s_R \to \tilde \tau \tilde \tau$ is
the main production process ($2 m_{\tilde \tau}< m_S < 2 m_{\tilde
  B}$), we should take into account the entropy production from the
$s_I$ decay which happens at a lower temperature. Therefore, for a
larger value of $r$, the stau abundance in the $m_S < 2 m_{\tilde B}$
region is more suppressed by a larger dilution effect.
For different values of the gravitino mass, the stau abundance
approximately scales as $Y_{\tilde \tau} \propto T_d^{-1} \propto
m_{3/2}$.

For the parameter set we took, $m_{\tilde \tau} =100$~GeV and
$m_{3/2}=30$~MeV, the staus decay rather early in the BBN era
($\tau_{\tilde \tau}=50$~sec), and thus there is no significant
constraint on $Y_{\tilde \tau}$ from BBN. The bound that the
gravitinos from the stau decays should not exceed the observed matter
energy density gives $Y_{\tilde \tau} \lesssim 1 \times 10^{-8}$,
which is satisfied for any value of $m_S$.
For $m_{3/2} = 100$~MeV and $m_{\tilde \tau}=100$~GeV, the stau lifetime
is 600~sec, with which we obtain a  BBN constraint from the D abundance,
$Y_{\tilde \tau} \lesssim 1 \times 10^{-10}$~\cite{Kawasaki:2008qe}. In this
case, a part of the parameter region is excluded as shown in the right
panel of Fig.~\ref{fig:Y-stau}.
For a further large value of $m_{3/2}$, the constraint from the $^6$Li
abundance becomes important. For example, for $m_{3/2} = 300$~MeV, the
constraint is $Y_{\tilde \tau} \lesssim 1 \times
10^{-13}$~\cite{Kawasaki:2008qe}, and the consistent parameter region
disappears for $m_{\tilde \tau} = 100$~GeV.

%%%%%%%%
\section{Non-thermal relic abundance}
\label{sec:NT} In this appendix we calculate the non-thermal relic
abundance of a particle $X$, assuming the following cosmological
scenario. (i) The energy density of the Universe is dominated by a
non-relativistic matter $\phi$ (e.g. a coherently oscillating scalar
field).  (ii) The $\phi$ field then decays into radiation and $X$, with
a decay rate $\Gamma_\phi$.\footnote{Note that this is different from
the case of Q-ball decay~\cite{FH}, in which the $X$ production suddenly
terminates at $T=T_d$.  The final $X$ abundance obtained here is about 5
times larger than the case of Q-ball decay.
} (iii) The subsequent pair annihilations of the $X$ particles reduce
its number until it freezes out.  The relevant Boltzmann equations are
given by (cf. \cite{Moroi:1999zb,Gelmini:2006pw})
\begin{eqnarray}
\dot{\rho_\phi} &=& -3H\rho_\phi - \Gamma_\phi \rho_\phi \,,
\label{eq:1}
\\
\dot{\rho}_{\rm rad} &=& -4H\rho_{\rm rad}+ \Gamma_\phi \rho_\phi \,,
\\
\dot{n}_X &=& -3H n_X - \langle \sigma v \rangle (n_X^2 - n_{X,eq}^2) + \Gamma_\phi \frac{\rho_\phi}{m_\phi} b \,,
\\
H^2 &=& \frac{1}{3 M_{\rm pl}^2} \rho_{\rm total},\qquad \rho_{\rm total} = \rho_\phi + \rho_{\rm rad}\,,
\label{eq:4}
\end{eqnarray}
where $\rho_\phi$ and $\rho_{\rm rad}$ are the energy density of the
$\phi$ and radiation, respectively, and we assume that the energy
density of the $X$ particle is negligible compared to them, $\rho_X \ll
\rho_{\rm total} = \rho_{\rm rad} + \rho_\phi$.  $H$ is the Hubble
parameter, $n_X$ is the number density of $X$, and $b$ is the averaged
number of $X$ particles produced per $\phi$.  Here and in what follows,
we assume that the equilibrium number density is negligible,
$n_{X,eq}\ll n_X$, which is a good approximation as long as $m_X \gg
T_d$ with a moderate value of $b$\footnote{Strictly speaking, $b$ must
satisfy $b \gg (m_\phi m_X^{3/2}/T_d^{5/2}) \exp(-m_X/T_d)$.
}, where $T_d$ is the decay temperature, defined by $T_d \equiv (\pi^2
g_*/90)^{-1/4}\sqrt{M_{\rm pl}\Gamma_\phi}$.  In terms of the
following variables,
\begin{eqnarray}
x \equiv \ln\lrf{\Gamma_\phi}{H}, \quad
f_\phi \equiv \frac{\rho_\phi}{\rho_{\rm total}}, \quad
N_X \equiv  \langle \sigma v \rangle \Gamma_\phi^{1/2} \frac{n_X}{H^{3/2}}\,.
\end{eqnarray}
the equations (\ref{eq:1})--(\ref{eq:4}) become 
\begin{eqnarray}
\left(1-\frac{f_\phi}{4}\right) \frac{d\,f_\phi}{d\,x}
&=& -\frac{1}{2}f_\phi\,e^x + \frac{1}{2} f_\phi (1- f_\phi) \,,
\label{eq:new1}
\\
\left(1-\frac{f_\phi}{4}\right) \frac{d\,N_X}{d\,x}
&=& -\frac{3}{8}f_\phi N_X
- \frac{N_X^2}{2\,e^{x/2}}
+ A f_\phi\,e^{x/2} \,.
\label{eq:new2}
\end{eqnarray}
This can be solved numerically with initial conditions $f_\phi(-\infty)=1$ and $N_X(-\infty)=0$,
and the final answer $N_X(\infty)$ depends only on the dimensionless parameter $A$, which is given by
\begin{eqnarray}
A &\equiv& \frac{3 M_{\rm pl}^2 \Gamma_\phi b \langle \sigma v \rangle}{2 m_\phi},\non\\
&\simeq& 7.7\times 10^6
\lrfp{g_*}{10}{\frac{1}{2}}
\lrfp{T_d}{100~{\rm MeV}}{2}
\lrf{\langle \sigma v \rangle}{10^{-7}~{\rm GeV}^{-2}}
\lrf{500~{\rm GeV}}{m_\phi}
\lrf{b}{1.0}\,.
\end{eqnarray}
The dependence of $N_X(\infty)$ on $A$ is actually very weak, and it
is empirically found that
\begin{eqnarray}
N_X(\infty) \simeq 4.5\left(1+0.043 \log \lrf{A}{10^6} \right)\,.
\end{eqnarray}
This approximation reproduces the numerical result within a few \%,
for a wide range of $A=10^3-10^{10}$.  Thus, the final $X$ abundance
(for $x\to\infty$, $f_\phi\to 0$) is given by, assuming $g_* =
const.$,
\begin{eqnarray}
\frac{n_X}{s} 
&=& N_X \frac{H^{3/2}}{\Gamma_\phi^{1/2} \langle \sigma v \rangle s}
= \lrfp{45}{8\pi^2 g_*}{1/2} \frac{N_X}{M_{\rm pl} T_d \langle \sigma v \rangle}
\\
&\simeq& 4.4\times 10^{-11}
\lrfp{10}{g_*}{1/2}
\lrf{100~{\rm MeV}}{T_d}
\lrf{10^{-7}~{\rm GeV}^{-2}}{\langle \sigma v \rangle}
\left(1+0.043 \log \lrf{A}{10^6} \right)\,.
\label{eq:NT}
\end{eqnarray}

\end{document}